\documentclass[a4paper,11pt]{article}
\pdfoutput=1
\usepackage{jheppub}
\usepackage[T1]{fontenc}
\usepackage{amsmath}
\usepackage{amssymb}
\usepackage{mathrsfs}
\usepackage{graphicx}
\usepackage{mathtools}
\usepackage{subcaption} 

\def\be{\begin{equation}}
\def\ee{\end{equation}}
\def\e{\text{e}}
\def\i{\text{i}}

\def\dd{\text{d}}

\def\max{\text{max}}

\def\AdS{\text{AdS}}
\def\V{\mathcal{V}}

\def\L{\mathcal{L}}

\begin{document}
\title{Linear Growth of Holographic Time-like Entanglement Entropy and Kasner exponents}
\author{Zi-Hao Li, Run-Qiu Yang}
\affiliation{Center for Joint Quantum Studies and Department of Physics, School of Science, Tianjin University, Yaguan Road 135, Jinnan District, 300350 Tianjin, P.~R.~China}
\emailAdd{lieaction.lzh@gmail.com}
\emailAdd{aqiu@tju.edu.cn}
\abstract{The holographic time-like entanglement entropy (TEE) extends entanglement to time-like boundary subregions. While its definitive holographic dictionary remains debated, one concrete proposal utilizes piece-wise extremal surfaces. In this work, we adopt this geometric prescription as an exploratory framework to holographically investigate the late-time ($\tau_0\to \infty$) growth of TEE in asymptotically AdS black holes with a space-like singularity and no inner horizon. By assuming a Kasner geometry near the space-like singularity and using the null energy condition, we analytically show that a critical extremal surface $\mathcal{A}_c$ inside the event horizon completely governs the late-time linear growth of the TEE. This result suggests that the late-time behavior of TEE is tightly constrained by the geometry of black hole interiors.  
While the dominant energy condition (DEC) guarantees an upper bound for the real part's growth rate, we conjecture a corresponding universal lower bound for the imaginary part. Numerical results from Einstein-scalar theory demonstrate the robustness of this bounding behavior: the vacuum Schwarzschild-AdS geometry consistently maximizes the real growth rate and minimizes the imaginary part, suggesting these bounds hold in broader holographic setups.}

\maketitle
\flushbottom 

\section{Introduction}\label{sec:intro}

The anti-de Sitter/conformal field theory (AdS/CFT) correspondence~\cite{Maldacena:1997re, Gubser:1998bc, Witten:1998qj} has provided a powerful framework for understanding how spacetime geometry is encoded in quantum entanglement of the boundary conformal field theory~\cite{Maldacena:2001kr, Maldacena:2013xja, VanRaamsdonk:2010pw}. A central example is the holographic duality between boundary entanglement entropy and bulk extremal surfaces. For static configurations, the Ryu-Takayanagi prescription computes the holographic entanglement entropy (HEE) of a spatial subregion from a codimension-2 space-like extremal surface anchored on the boundary~\cite{Ryu:2006bv, Hubeny:2007xt}.

In time-dependent settings, the evolution of entanglement entropy has also been studied extensively. From the viewpoint of quantum information, its growth rate helps us understand how correlations spread after a quench and how a many-body system approaches a highly entangled state~\cite{Calabrese:2005in, Ho:2015woa}.
In the usual holographic study of entanglement growth, such as the Hartman-Maldacena setup~\cite{Hartman:2013qma, Li:2022cvm}, the entropy of a fixed spatial region evolves with the boundary time as an external parameter. Since spacetime treats space and time on the same footing, there is strong motivation to explore the time-like entanglement entropy (TEE), $S_\mathcal{T}$, as a novel quantum information measure that characterizes temporal entanglement.

Following early proposals for time-like entanglement~\cite{Olson:2011bq, Wang:2018jva}, Ref.~\cite{Doi:2022iyj} introduced a concrete definition of TEE from both boundary and bulk viewpoints. Holographically, TEE generalizes the idea of HEE to \textit{time-like} separated regions and at present, the precise holographic dictionary for TEE remains an open question. On the boundary field theory side, several approaches have been developed, including operator-algebraic definitions~\cite{Jiang:2025pen}, real-time Schwinger-Keldysh or replica methods~\cite{Milekhin:2025ycm, Gong:2025pnu, Guo:2025ase, Guo:2024lrr, Xu:2024yvf}, non-Hermitian density matrix~\cite{Harper:2025lav}, AdS/BCFT constructions~\cite{Chu:2023zah}, gravitational anomalies~\cite{Chu:2025sjv}, and as diagnostics for quantum chaos and ergodicity~\cite{Das:2025fcd, Das:2026ifj}. Complementary to these boundary perspectives, various bulk constructions have been proposed, such as top-down string models~\cite{Nunez:2025gxq, Nunez:2025puk}, modular flow methods~\cite{He:2023ubi, Wen:2024yny}, non-relativistic gravity and Lifshitz setups~\cite{Afrasiar:2024lsi, Afrasiar:2024ldn, Afrasiar:2025eam, Prihadi:2026nua, Jena:2024tly}, extremal surfaces associated with pseudo-entropy in de Sitter space~\cite{Narayan:2022afv, Narayan:2023zen, Narayan:2023ebn, Takayanagi:2025ula}, and complex bulk extremal surfaces~\cite{Heller:2024whi, Guo:2025pru, Heller:2025kvp}. 
 
Among these proposals, the Complex-valued Weak Extremal Surface (CWES) prescription~\cite{Doi:2022iyj, Li:2022tsv, Doi:2023zaf, Bohra:2025mhb, Li:2026fcr} provides a concrete and tractable exploratory framework. In this prescription, the geometric dual is built from piece-wise space-like and time-like bulk segments. While we do not claim to favour the CWES over other proposals, nor that it is the definitively correct holographic dual for TEE, it is especially useful as a tool to investigate the structural insights that a holographic window into time-like entanglement can reveal concerning the relation between the deep black hole interior and the TEE growth rate.

\begin{figure}[htbp]
 \begin{center}
   \includegraphics[width=0.72\textwidth]{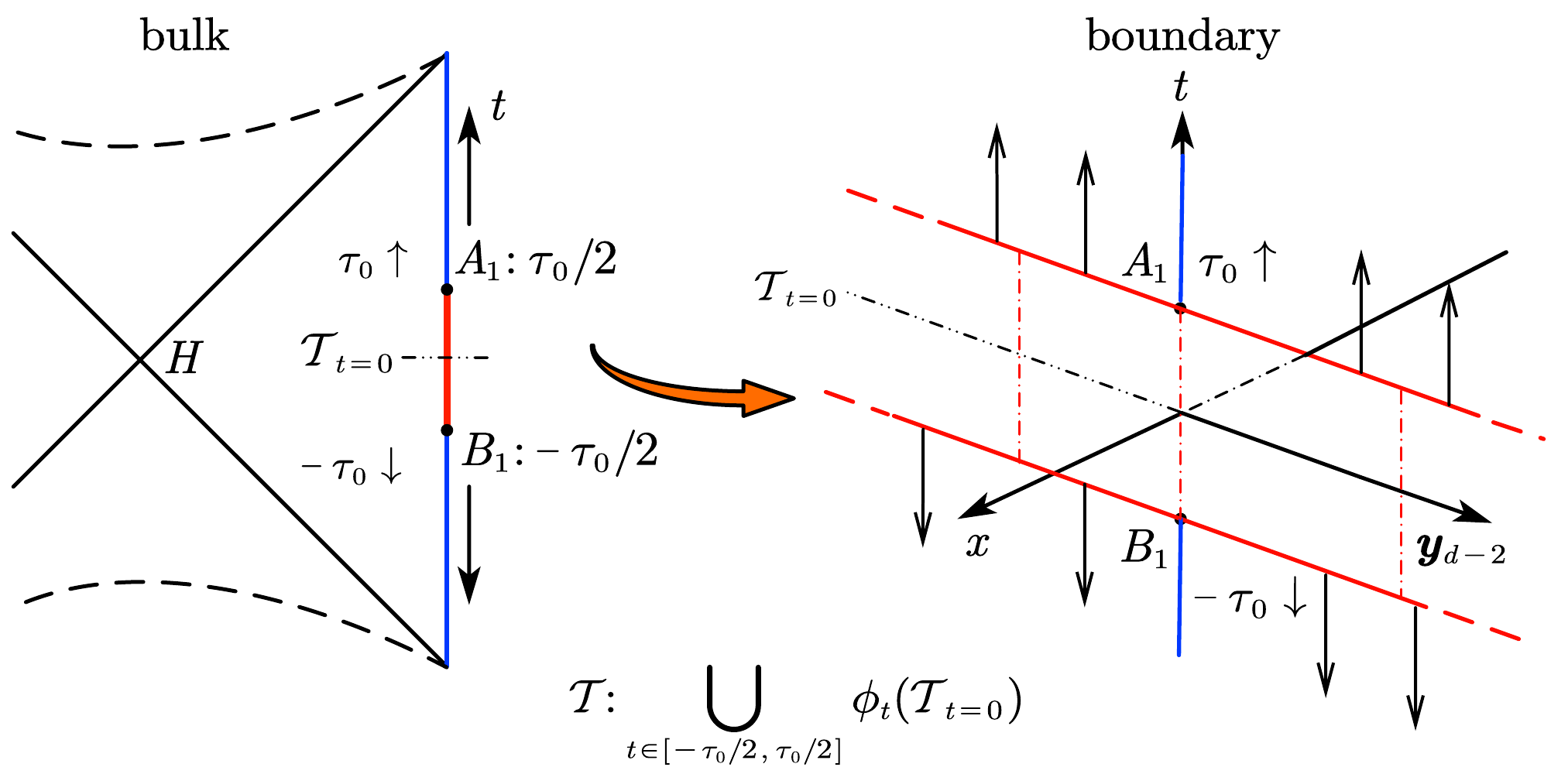}
\end{center}
\caption{Schematic diagram illustrating the growth of TEE. The time-like strip (red region) $\mathcal{T}$ lies on the asymptotic AdS boundary. The AdS boundary is spanned by coordinates $\{t,x,\boldsymbol{y}_{d-2}\}$, where $\boldsymbol{y}_{d-2}$ stands for $(d-2)$-dimensional coordinates $\{y_1,y_2,\cdots,y_{d-2}\}$. The strip is defined at the AdS boundary with $x=0$ and $t\in[-\tau_0/2,\tau_0/2]$. As the width of the strip $\tau_0$ increases, the time-like subregion encompasses more degrees of freedom that originally belonged to the future region (blue region) and the corresponding TEE changes accordingly.} \label{fig:time-evolution}
\end{figure}

With the CWES framework as our starting point, we investigate the growth rate of TEE in a static spacetime background. We consider a $(d-1)$-dimensional time-like strip $\mathcal{T}$ on the boundary. This subregion is defined by the dynamical time evolution of a fixed spatial cross-section. Let $\mathcal{T}_{t=0}$ be a $(d-2)$-dimensional space-like surface at $t=0$ (specifically, located at $x=0$ and spanning the transverse spatial directions, see Fig.~\ref{fig:time-evolution}).
Using the time evolution operator $\phi_t$, the symmetric time-like strip is generated by the union of these evolved slices over a temporal width $\tau_0$: $\mathcal{T} = \bigcup_{t \in [-\tau_0/2,\, \tau_0/2]} \phi_t(\mathcal{T}_{t=0})$. As shown in Fig.~\ref{fig:time-evolution}, $\mathcal{T}$ consists of a continuous sequence of temporal slices representing the past and future states of $\mathcal{T}_{t=0}$.
Conceptually, the TEE may be interpreted as a probe of intrinsic temporal quantum correlations among these successive futures (and pasts) of $\mathcal{T}_{t=0}$ contained within the set $\mathcal{T}$. As we dynamically expand this temporal observation window, the width $\tau_0$ symmetrically increases. Consequently, the expanding subregion $\mathcal{T}$ includes an increasing number of $\mathcal{T}_{t=0}$'s future (and past) slices.
Therefore, the growth rate of TEE $\dd S_{\mathcal{T}} / \dd \tau_0$ measures the entanglement strength between the existing temporal history and the newly included future slices. This physical picture is fundamentally different from the familiar HEE growth in the Hartman-Maldacena (HM) setup for a thermofield double (TFD) state~\cite{Hartman:2013qma, Li:2022cvm}, where entropy grows because a boundary spatial subregion of a fixed size is globally translated along the time parameter.

While previous studies have explored the dynamics of time entanglement~\cite{Liu:2022ugc} and related concepts, such as pseudo-entropy after a quantum quench in lower-dimensional quantum systems~\cite{Guo:2022sfl, Omidi:2023env, He:2023syy, Narayan:2023ebn}, higher-dimensional field theories~\cite{Mukherjee:2022jac}, and AdS$_3$-Vaidya spacetimes~\cite{Katoch:2025bnh, Katoch:2026dzs}, a comprehensive understanding of the TEE growth rate and its potential universal bounds remains incomplete.
In this paper, we utilize the CWES framework to systematically investigate the dynamical growth of TEE in planar symmetric asymptotic AdS black holes featuring only space-like singularities (i.e., lacking inner horizons).
Using asymptotic integrals and the Hamilton-Jacobi formalism, we analytically prove that this growth rate is determined by the local geometry at an interior\footnote{In addition to the cases of space-like entanglement entropy~\cite{Hartman:2013qma,Li:2022cvm} and our present result, similar critical-surface behavior has also appeared in studies of complexity~\cite{Carmi:2017jqz, Yang:2019alh, An:2022lvo, Auzzi:2022bfd} (for the late-time evolution of holographic complexity in cosmological models, see also~\cite{Paul:2025gpk}) and in other TEE proposals~\cite{Heller:2024whi, Heller:2025kvp}.} \textit{critical extremal surface} $\mathcal{A}_c$. Furthermore, we find that the universal existence of this critical surface is guaranteed by the null energy condition (NEC) in the bulk. This finding provides a novel manifestation of the NEC's fundamental importance in holographic duality: in addition to ensuring the monotonicity of renormalization group flows (the holographic c-theorem)~\cite{Freedman:1999gp, Myers:2010xs} and the strong sub-additivity of spatial entanglement~\cite{Headrick:2007km, Wall:2012uf}, the bulk NEC also protects the well-defined late-time linear growth of time-like entanglement. Moreover, we reveal that this geometric protection has a finite tolerance, ensuring that the linear growth rate remains robust even when the classical NEC is mildly compromised by expected quantum gravity effects in spacetimes with Kasner-like singularities.

Motivated by space-like entanglement studies demonstrating that the vacuum black hole maximizes the growth rate at a fixed entropy density~\cite{Hartman:2013qma, Li:2022cvm}, we investigate whether an analogous upper bound holds for the late-time growth rate of the real part of the TEE. In Appendix~\ref{appendix:DEC}, we prove that this rate is indeed maximized by the Schwarzschild-AdS geometry under the dominant energy condition. Furthermore, we conjecture that this vacuum background simultaneously provides a universal lower bound for the imaginary part of the TEE. To test the robustness of these bounds, we numerically construct hairy black holes by introducing a real scalar field coupled to gravity.  We systematically evaluate the TEE across thermodynamic ensembles with fixed horizon radius, fixed temperature, and fixed total energy density. Our results reveal that in both the fixed horizon radius and fixed temperature ensembles, the pure vacuum geometry consistently maximizes the real growth rate and minimizes the imaginary part.
Crucially, however, while these bounds appear to be explicitly violated under alternative quantization schemes when the total energy density is constrained, we demonstrate that they are recovered by utilizing a newly proposed scheme-independent thermodynamic formulation~\cite{Li:2026dgx}.  

The remainder of this paper is organized as follows. In Sec.~\ref{sec:GrowthTEE}, we compute the TEE for the above black hole backgrounds and employ two distinct methods to prove the linear growth rate of TEE.  By analyzing the asymptotic Kasner behavior near the singularity and using NEC, we analytically show the existence of a critical extremal surface $\mathcal{A}_c$ that universally governs this late-time $\tau_0\to \infty$ growth. In Sec.~\ref{sec:ImagofTEE}, we numerically investigate the upper bound of the real part's growth rate and the lower bound of the imaginary part within a hairy black hole model.
Finally, Sec.~\ref{sec:discussion} summarizes our findings and outlines open questions. Additional technical details are provided in the appendices: Appendix~\ref{appendix:threecase} classifies alternative dynamical configurations for the time-like boundary strip (specifically comparing temporal width expansion, semi-infinite TFD setups, and global rigid time translations). Appendix~\ref{appendix:DEC} derives a DEC-based comparison theorem for interior growth potentials, applies it to the real TEE growth rate.

\section{Complex Area of Time-like Entanglement and Its Growth Rate}\label{sec:GrowthTEE}

In this section, we first introduce the holographic calculation of time-like entanglement entropy (TEE) in an asymptotically AdS black hole spacetime. We then present two distinct methods for the late-time linear growth of the real part and prove the existence of the critical extremal surface that controls this late-time growth.

\subsection{Setup}\label{subsec:setup}

We consider a $D=(d+1)$-dimensional static asymptotic AdS black hole, with spherical, planar or hyperbolic symmetries. The bulk metric reads
\be\label{eq:AdSmetric}
    \dd s^2  =\frac{1}{z^2}\left(-f(z)\e^{-\chi(z)}\dd t^2+f^{-1}(z)\dd z^2+\dd \Sigma^{2}_{k,d-1}\right)\ .
\ee
Here, $k \in \{+1, 0, -1\}$ describes the spatial curvature of the ($d-1$)-dimensional section $\dd \Sigma^{2}_{k,d-1}$, given by
\be\label{eq:k-metric}
\dd \Sigma^{2}_{k,d-1} =\begin{dcases}
 \dd\theta^2+\sin^2\theta\,\dd\Omega^2_{d-2}\ ,  & \text{for}\, k=+1\ ; \\
  \dd x^2+\dd \boldsymbol{y}_{d-2}^{2}\ , & \text{for}\, k=0\ ;\\
  \dd\sigma^2+\sinh^2\sigma\,\dd\Omega^2_{d-2}\ ,& \text{for}\, k=-1\ .
 \end{dcases}
\ee
For simplicity and direct relevance to holography, we focus on the planar case ($k=0$) and set $G_\text{N}=1$, $\ell_\AdS=1$ throughout this paper.

To compute the TEE for a specific subregion in the dual boundary CFT$_d$, we employ the Complex-valued Weak Extremal Surface (CWES) prescription~\cite{Doi:2022iyj, Li:2022tsv, Doi:2023zaf, Li:2026fcr}. This prescription instructs us to find a special piece-wise smooth codimension-2 bulk surface $\Gamma_{\mathcal{T}}$ anchored to the boundary time-like subregion of interest, $\mathcal{T}$. The TEE is holographically computed by evaluating the complex-valued area of this minimal extremal surface:
\be\label{eq:CWESofTEE}
S_\mathcal{T}=\text{Min}\left \{\mathop{\text{Ext}}\limits_{\partial \Gamma_\mathcal{T}}\frac{\mathscr{A}(\Gamma_\mathcal{T})}{4G_\text{N}^{(d+1)}}\right \}\ .
\ee
Here, $\mathscr{A}(\Gamma_{\mathcal{T}})$ denotes the complex-valued area of the piece-wise smooth surface $\Gamma_{\mathcal{T}}$. The operations ``Ext'' and  ``Min'' signify that we must find a surface that is both extremal (its area is stationary under local deformations) and minimal among all such valid candidates, according to the specific ordering relation ``$\prec$'' (or ``$\succ$'')  defined by~\cite{Li:2022tsv}.

\begin{figure}[htbp]
 \begin{center}
   \includegraphics[width=0.5\textwidth]{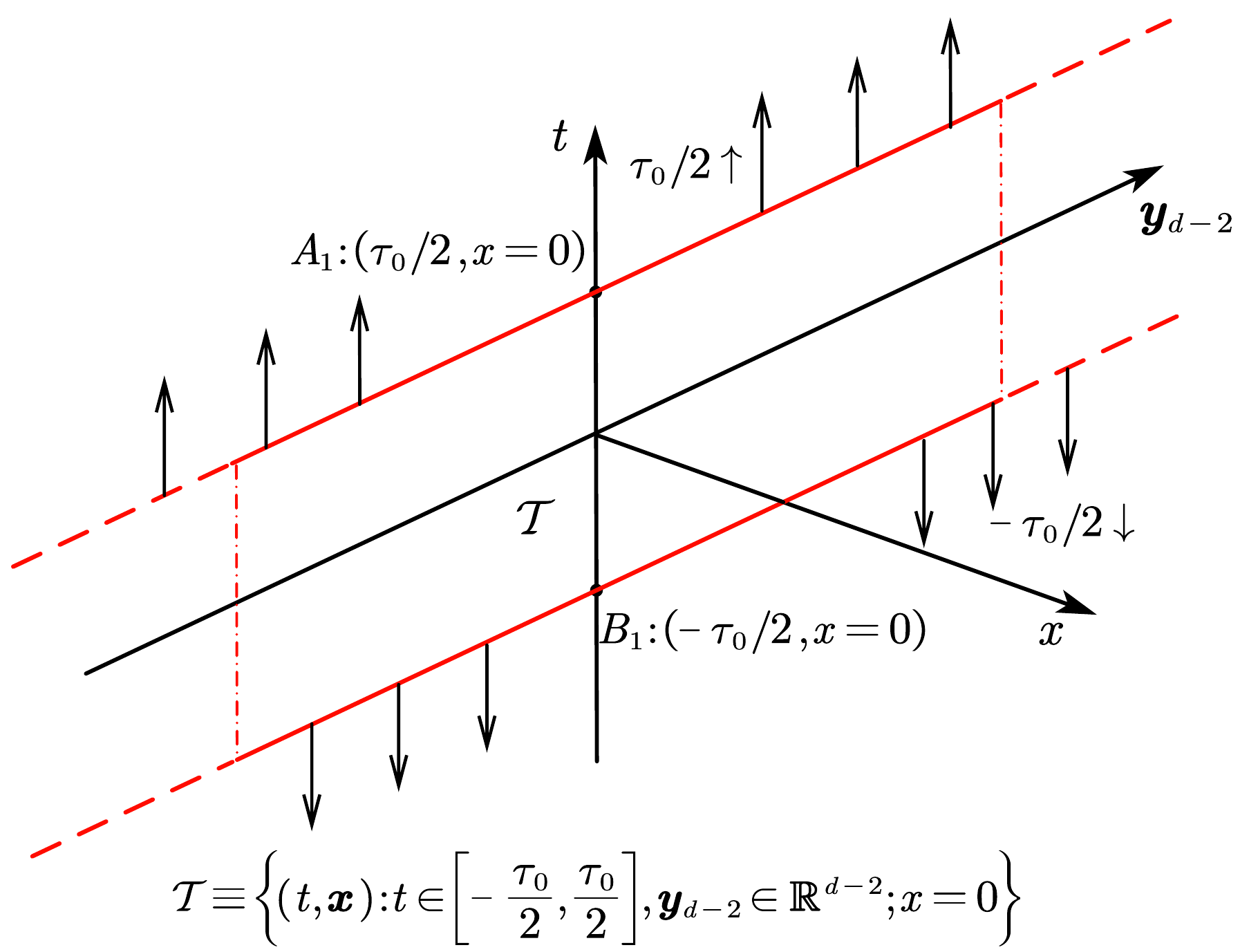}
\end{center}
\caption{A schematic diagram of a time-like strip (depicted in red) on the boundary. As the temporal width $\tau_0$ of the strip uniformly increases, the strip includes more degrees of freedom from its future complement.} \label{fig:time-like-strip}
\end{figure}

As depicted in Fig.~\ref{fig:time-like-strip}, the boundary subregion $\mathcal{T} = \bigcup_{t \in [-\tau_0/2,\, \tau_0/2]} \phi_t(\mathcal{T}_{t=0})$ we consider is a time-like ``strip'' extending infinitely in the $d-2$ spatial directions and possessing a temporal width $\tau_0$ symmetric around $t=0$:
\be\label{eq:time-like strip}
\mathcal{T} \equiv \left\{ (t,\boldsymbol{x}):t\in \left[ -\frac{\tau_0}{2},\frac{\tau_0}{2} \right] ,\boldsymbol{y}_{d-2}\in \mathbb{R}^{d-2};x=0 \right\}\ .
\ee
The two boundary components of time-like strip $\mathcal{T}$ on boundary CFT$_d$ are denoted by $A_1$ and $B_1$ with the time coordinate $t = \pm \frac{\tau_0}{2}$. Unlike the usual spatial HM setup based on a thermofield double state, from the field theory perspective, this time-like strip selects degrees of freedom in time direction in a single-sided CFT.
As indicated by the arrows in Fig.~\ref{fig:time-like-strip}, we track the change in the minimal CWES area as the temporal width $\tau_0$ increases\footnote{Note that there are two other alternative configurations for evaluating the TEE growth rate, detailed in Appendix~\ref{appendix:threecase}.}. Holographically, as depicted from a bulk perspective in Fig.~\ref{fig:late-time}(a), this geometric variation captures the time evolution of the time-like entanglement entropy on the boundary. Within the CWES framework, this dynamical process serves as a potential holographic dual candidate to characterize the growth of temporal quantum entanglement within the boundary CFT$_d$.

\begin{figure}[htbp]
 \begin{center}
   \includegraphics[width=0.8\textwidth]{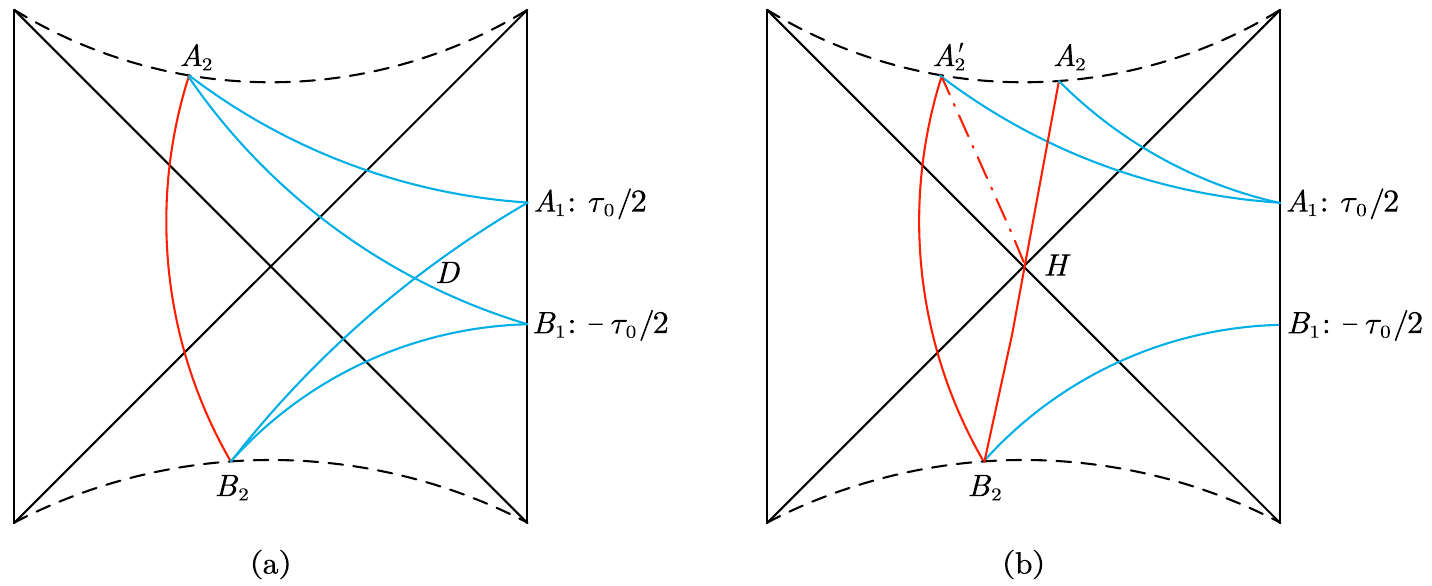}
 \end{center}
\caption{Configurations of CWES of time-like strip in an asymptotic black hole. Blue lines stand for space-like surface and green lines represent time-like surface.  (a) There are two types of relevant configurations of CWESs, which are $A_1A_2B_2B_1$ and $A_1B_2A_2B_1$.  (b) Comparison of time-like paths: The symmetric surface $A_2B_2$ passes through the bifurcation surface $H$ of the event horizon, whereas $A'_2B_2$ represents a generic path.}  \label{fig:CWES-AdS-Sch}
\end{figure}

To explicitly perform this calculation, we must identify the correct minimal CWES configuration among all candidates that penetrate the black hole interior (further rigorous proofs can be found in Refs.~\cite{Li:2022tsv, Li:2026fcr}). First, we define the complex-valued area density as $\mathcal{A}=\mathscr{A}/\mathcal{V}_{d-2}$, where $\mathcal{V}_{d-2}\equiv\int\dd^{d-2}y$ is the infinite transverse volume.
As shown in Fig.~\ref{fig:CWES-AdS-Sch}(a), there are two relevant piece-smooth configurations: $A_1A_2B_2B_1$ and $A_1B_2A_2B_1$. By applying arguments analogous to the ``triangle inequality'' for space-like surfaces, one can show that $\mathcal{A}(A_1A_2B_2B_1) \prec \mathcal{A}(A_1B_2A_2B_1)$. Thus, the configuration $A_1A_2B_2B_1$ dominates.

Next, we determine the exact path of the time-like segment. As illustrated in Fig.~\ref{fig:CWES-AdS-Sch}(b), we compare a generic ``non-crossing'' path $A'_2B_2$ with a ``crossing'' path $A_2B_2$ that explicitly passes through the bifurcation surface $H$. Since time-like surfaces obey the ``anti-triangle inequality'', a generic path satisfies $\mathcal{A}(A_2'B_2) \succ \mathcal{A}(A_2'H) + \mathcal{A}(HB_2)$. Due to the time-translation symmetry of the static AdS background, $\mathcal{A}(A_2'H) = \mathcal{A}(A_2H)$. Consequently, we obtain:
\begin{equation}
     \mathcal{A}(A_1A_2B_2B_1)\prec \mathcal{A}(A_1A'_2B_2B_1)\ .
\end{equation}
This shows that it is enough to restrict attention to configurations whose time-like segment passes through the bifurcation surface.

\begin{figure}[htbp]
 \begin{center}
   \includegraphics[width=0.8\textwidth]{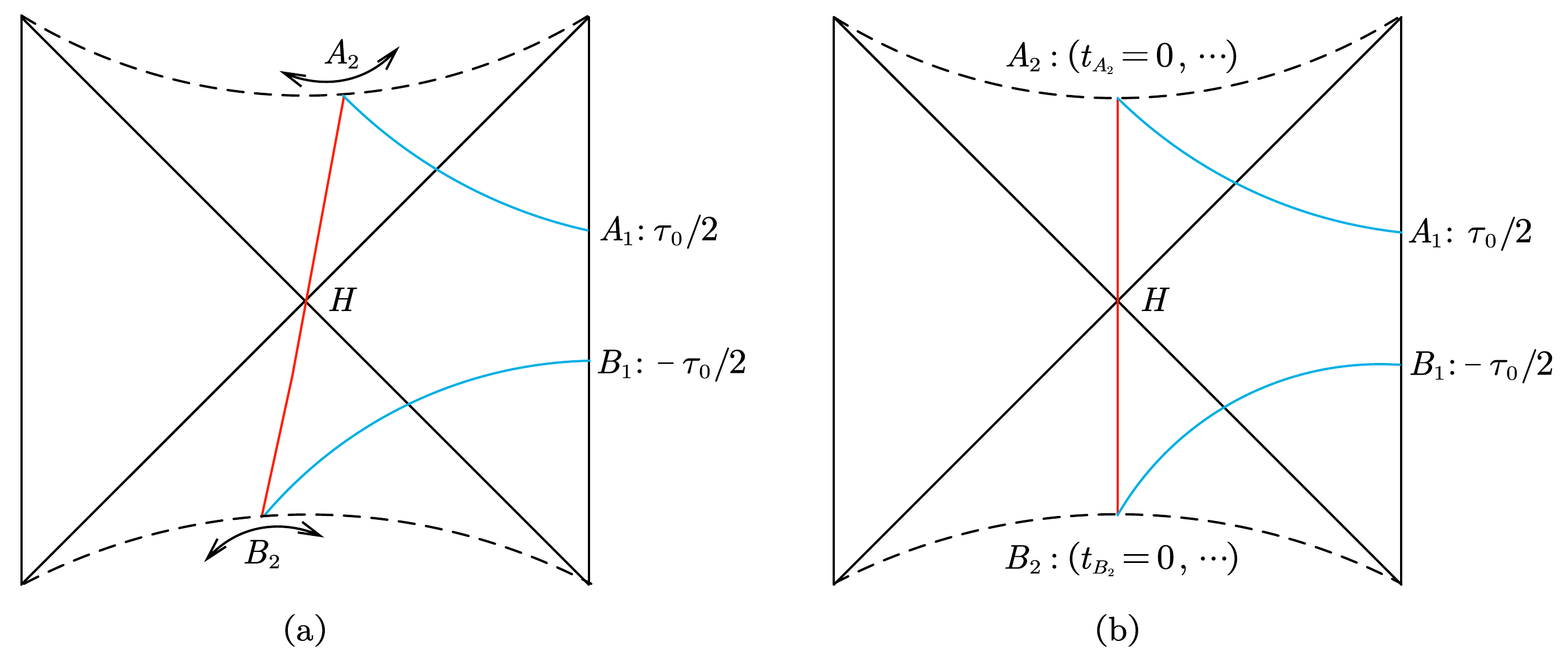}
 \end{center}
\caption{(a) The real part of the CWES area density plotted as a function of the bulk location of the junction point $A_2$. Smoothly varying this position generates a family of candidate extremal surfaces. (b) The minimal extremal configuration of CWES, corresponding to the time-symmetric case $t|_{A_2}=t|_{B_2}=0$. } \label{fig:Hamilton-Jacobi}
\end{figure}

Finally, as  proven in Ref.~\cite{Li:2026fcr} using the Hamilton-Jacobi method, among all these valid candidates, the time-symmetric ``vertical'' configuration ($t|_{A_2}=t|_{B_2}=0$, shown in Fig.~\ref{fig:Hamilton-Jacobi}(b)) inherently corresponds to the minimal CWES.
This symmetry significantly simplifies the total TEE calculation: we only need to evaluate the area of the single space-like branch $A_1A_2$, double its value to account for both sides, and combine it with the purely imaginary contribution from the time-like segment $A_2B_2$:
\begin{equation}
    \mathcal{A}_{A_1A_2B_2B_1}(\tau_0)  =2\text{Re} \mathcal{A}_{A_1A_2}(\tau_0) +\i\,\text{Im} \mathcal{A}_{A_2B_2}\ .
\end{equation}

\subsection{Real and Imaginary Parts of Time-like Entanglement}\label{subsec:R-IofTEE}

We now proceed to explicitly evaluate the area of the minimal CWES identified in the previous subsection.

\subsubsection{Real part}

To properly describe the space-like segment $A_1 A_2$ that smoothly crosses the event horizon and terminates at the singularity, we must avoid the coordinate singularity at the horizon $z=z_h$. We therefore switch to the ingoing Eddington coordinate patch:
\begin{equation}\label{eq:Eddington0}
        v = t-z^*(z)\ , \quad \text{where} \quad z^*(z)=\int_0^z{\dd \tilde{z}\,f^{-1}(\tilde{z})\e^{\chi (\tilde{z})/2}}\  .
\end{equation}
In this coordinate system, the metric behaves regularly across the horizon and becomes:
\begin{equation}\label{eq:Eddington}
        \dd s^2  = \frac{1}{z^2}\left( -f(z)\e^{-\chi (z)}\dd v^2-2\e^{-\chi(z) /2}\dd v\dd z+\dd x^2+\dd \boldsymbol{y}_{d-2}^{2} \right)\ .
\end{equation}

Using the Eddington metric~\eqref{eq:Eddington}, we parameterize the codimension-2 space-like surface $A_1A_2$ by the coordinates $\{v, z=z(v), x=0, \boldsymbol{y}_{d-2} \in \mathbb{R}^{d-2}\}$.
The area density functional $\mathcal{A} = \mathscr{A}/\mathcal{V}_{d-2}$ for surface $A_1A_2$ is given by
\begin{equation}\label{eq:Lagrangian}
    \begin{aligned}
        \text{Re}\mathcal{A}_{A_1 A_2} &=\int \dd v\,L\ ,\\
        L(z, z') &= z^{1-d} \sqrt{-f(z) \e^{-\chi (z)}-2\e^{-\chi (z) /2}z'}\ ,
    \end{aligned}
\end{equation}
where $L$ is the Lagrangian, and $z' \equiv \frac{\dd z}{\dd v}$.

Since the static spacetime background ensures that the Lagrangian $L$ has no explicit dependence on $v$, the associated Hamiltonian $E=z^{\prime}\frac{\partial L}{\partial z^{\prime}}- L$ is a conserved quantity along the extremal surface:
\begin{equation}\label{eq:conservedH}
    \begin{aligned}
        E &= \frac{z^{1-d}\left( f\e^{-\chi}+\e^{-\chi /2}z^{\prime} \right)}{\sqrt{-f\e^{-\chi}-2\e^{-\chi /2}z^{\prime}}}\\
        &\equiv -H_*^2\ .
    \end{aligned}
\end{equation}
Here, we denote this positive real constant by $H_*^2$. By solving\footnote{Directly solving for $z^\prime$ from Eq.~\eqref{eq:conservedH} leads to $z^\prime = - \e^{\chi /2} X (X \pm H_{*}^{2}z^{d-1})$. Here, only the positive branch ``$+$'' is admissible because the negative sign would imply $z' = 0$ at the horizon, which is not physically allowed.} the equation of motion~\eqref{eq:conservedH} for $z'$ in terms of $z$ and $H_*$, we obtain the differential equation governing the surface $A_1 A_2$ :
\begin{equation}\label{eq:zprime}
        z'(v) = - \e^{\chi /2} X (X + H_{*}^{2}z^{d-1})\ ,
\end{equation}
where we have introduced the auxiliary function $X$ as $X(H_*) = \sqrt{H_{*}^{4}z^{2d-2}+f(z)\e^{-\chi(z)}}$.

For the time-symmetric minimal configuration discussed previously, the coordinate $z$ runs from $0$ (the boundary) to $\infty$ (the singularity) as $v$ runs from $\tau_0/2$ to $0$ (since $v_{A_2}=0$).  Applying these boundary conditions to the endpoints, $A_1: (z=0, v=\tau_0/2)$ and $A_2: (z=\infty, v=0)$, we can express the boundary temporal width $\tau_0$ as:
\begin{equation}\label{eq:tau_0&H_*}
    \tau_0(H_*) = 2\int_0^\infty \frac{\dd z}{\e^{\chi /2} X (X + H_{*}^{2}z^{d-1})}\ .
\end{equation}
Similarly, the total real area density for the TEE is given by $2\text{Re}\mathcal{A}_{A_1A_2} = 2\int \dd v\,L$. To evaluate the total real area density $2\text{Re}\mathcal{A}_{A_1A_2}$, we must carefully isolate the UV divergence near the asymptotic AdS boundary $z \to 0$. By subtracting\footnote{For $d=2$, the subtraction is logarithmic and should be treated separately~\cite{Li:2026fcr}, in this paper we mainly focus on $d>2$.} the purely divergent vacuum contribution~\cite{Ryu:2006bv, Li:2026fcr}, we obtain the finite, regularized real area directly:
\begin{equation}\label{eq:A&H_*}
        \text{Re}\mathcal{A}_{\text{reg}}(H_*) = 2\int_0^\infty \dd z \left( \frac{z^{1-d}}{\e^{\chi /2} X} - z^{1-d} \right)\ .
\end{equation}
Here we subtracted the UV vacuum divergent $2\int_\epsilon^\infty \dd z\,z^{1-d}\equiv \frac{2\epsilon^{2-d}}{d-2}$, where $\epsilon$ is the UV cutoff.
Equations~\eqref{eq:tau_0&H_*} and~\eqref{eq:A&H_*} form a parametric system. By systematically varying the constant $H_*$, we can numerically generate a sequence of paired values $(\tau_0(H_*), \text{Re}\mathcal{A}_{\text{reg}}(H_*))$. Plotting these pairs directly yields the functional dependence of the real part of TEE on the temporal width, $\text{Re}\mathcal{A}_{\text{reg}}(\tau_0)$, effectively bypassing the need to analytically invert the complicated parametric integrals:
\be
    \begin{dcases}
        \tau_0  = \tau_0(H_*)\ ,\\
        \text{Re}\mathcal{A}_{\text{reg}}  =\text{Re}\mathcal{A}_{\text{reg}}(H_*)\ ;
    \end{dcases}
        \ \Longleftrightarrow\ \text{Re}\mathcal{A}_{\text{reg}}=\text{Re}\mathcal{A}_{\text{reg}}(\tau_0)\ .
\ee

\subsubsection{Imaginary part}\label{subsubsec:imagipart}

The imaginary part of the TEE arises from the area of the time-like segment $A_2 B_2$, which connects the past and future singularities. As established in Sec.~\ref{subsec:setup}, for the minimal configuration, this segment passes through the bifurcation surface $H$, as shown in Fig.~\ref{fig:Hamilton-Jacobi}(b). To calculate its area, we evaluate the area of a surface at constant $t$ and $x$, extending from the singularity $z=\infty$ to the horizon $z=z_h$ and back. The induced metric on this codimension-2 time-like surface is
\be
\begin{aligned}
    \dd s^2_\text{TL} &=\frac{1}{z^2}\left[ f^{-1}(z)\dd z^2+\dd \boldsymbol{y}_{d-2}^{2} \right]\ ,
\end{aligned}
\ee
Since this segment resides entirely inside the horizon ($z>z_h$), the blackening factor $f(z)$ is strictly negative. Consequently, the area element is $\dd\,\text{Im}\mathscr{A} = \sqrt{|h|} \,\dd z \,\dd^{d-2}y = z^{1-d} / \sqrt{-f(z)} \,\dd z \,\dd^{d-2}y$. Accounting for both the past and future interior regions, the total imaginary area density $\text{Im} \mathcal{A} \equiv \text{Im}\mathscr{A}/\mathcal{V}_{d-2}$ is twice this integral:
\be\label{eq:imaginarypart}
\text{Im} \mathcal{A} \left( A_1B_1B_2A_2 \right) =2\int_{z_h}^{\infty}{\dd z\,\frac{z^{1-d}}{\sqrt{-f\left( z \right)}}} \ .
\ee

This finite, purely imaginary component possesses several unique physical properties. First, unlike the real part, it does not suffer from UV divergences; it is a purely infrared (IR) quantity localized within the black hole interior. Second, as explicitly shown in Eq.~\eqref{eq:imaginarypart}, it is entirely independent of the boundary temporal width $\tau_0$ and conserved quantity $H_*$. Instead, it is fixed by the internal metric function $f(z)$, and may therefore encode physical information about the black hole's temperature and interior degrees of freedom~\cite{Li:2026fcr}, which is not directly accessed by standard space-like HEE surfaces. Because it is highly sensitive to the deep interior geometry, this imaginary part provides a theoretical probe to test whether geometric bounds exist for time-like entanglement, a question we systematically investigate in Section~\ref{sec:ImagofTEE}.

\subsection{Late-time Linear Growth of Time-like Entanglement}\label{subsec:asymptau}

In this subsection, we demonstrate that the TEE exhibits a linear growth phenomenon at late times $\tau_0\to \infty$. As established in Sec.~\ref{subsec:R-IofTEE}, numerically varying the conserved quantity $H_*$ captures the dynamical evolution of the TEE.
However, it is crucial to recognize that the conserved quantity $H_*$ cannot be arbitrarily small; it is  bounded by a physical minimum $H_{*,\min}$ to ensure the bulk extremal surface remains well-defined. As we will explicitly show below, taking the late-time limit ($\tau_0 \to \infty$) on the boundary is mathematically equivalent to pushing the conserved quantity to this lower bound ($H_* \to H_{*,\min}$). We employ two distinct analytical methods to evaluate this late-time growth rate, both confirming the existence of a universal bound governed by a critical bulk surface $\mathcal{A}_c$.

\subsubsection{Method 1: Asymptotic Analysis of the Area Integral}

\begin{figure}[htbp]
 \begin{center}
   \includegraphics[width=0.99\textwidth]{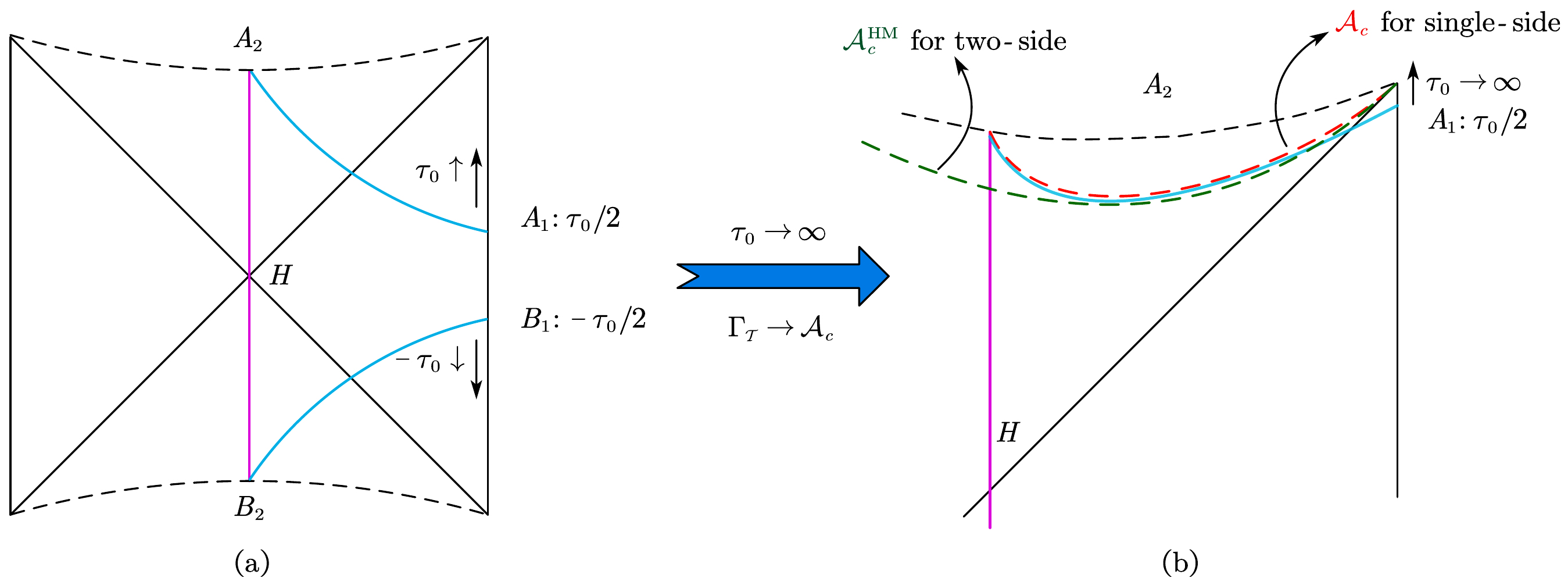}
 \end{center}
\caption{Schematic diagrams illustrating the time evolution and late-time limit of the TEE configurations. (a) The initial time evolution of the CWES. (b) The late-time limit as $\tau_0 \to \infty$. The space-like part of the CWES (solid blue line) asymptotically approaches the critical extremal surface $\mathcal{A}_c$ (red dashed line) located at $z=z_c$.
For comparison, the green dashed line represents the Hartman-Maldacena (HM) surface. In the late-time limit, the dynamic TEE surface tightly hugs both its own critical extremal surface $\mathcal{A}_c$ and the HM surface's critical region.} \label{fig:late-time}
\end{figure}

To ensure that the area functional~\eqref{eq:A&H_*} remains real-valued along the entire bulk trajectory, the quantity inside the square root of the auxiliary function $X(z)$ must be strictly non-negative. This imposes a fundamental geometric constraint on the conserved quantity $H_*$:
\begin{equation}
    H_{*}^{4}z^{2d-2}+f(z)\e^{-\chi(z)} \geqslant 0 \quad \implies \quad H_{*}^{4} \geqslant -z^{2-2d}f(z)\e^{-\chi(z)}\ .
\end{equation}
To simplify the analysis, we define a function $g(z) \equiv -z^{2-2d}f(z)\e^{-\chi(z)}$. For a valid real solution to exist for all $z$ along the surface, $H_*^4$ must be greater than or equal to the global maximum of $g(z)$ along the physical path:
\begin{equation}\label{eq:Hstar}
    H_{*}^{4} \geqslant \max_{z>z_h}\{g(z)\}\ .
\end{equation}
This inequality defines the minimal allowed value $H_{*,\min}$.  Geometrically, as depicted in Fig.~\ref{fig:late-time}(b), it identifies a \textit{critical extremal surface}, denoted as $\mathcal{A}_c$, located at a specific radial position $z=z_c$ where this maximum is achieved:
\begin{equation}\label{eq:Hstarmin}
    H_{*,\min}^4 \equiv g(z_c)\ , \quad \text{where} \quad g'(z_c) = 0\ .
\end{equation}
If $g(z)$ has multiple local maxima, the minimal allowed value $H_{*,\min}^4$ is determined by the global maximum. As $H_* \to H_{*,\min}$, the bulk space-like surface directly passes through any smaller local maxima and asymptotically approaches the critical surface located at the global maximum. If there are multiple global maxima, say $\{z_{c,1}, z_{c,2}, z_{c,3}, \dots\}$ and $z_{c,1} < z_{c,2} < z_{c,3} < \dots$ with the same value, the surface will approach the outermost one, i.e, $\mathcal{A}_{c,1}$ located at $z_{c,1}$, as shown in Fig.~\ref{fig:late-time}(b).
To ensure this dynamical mechanism is physically viable, we must confirm that such a critical surface $\mathcal{A}_c$ indeed exists and resides  within the black hole interior, satisfying $z_h < z_c < \infty$. In the next subsection (Sec.~\ref{subsec:existence}), we will use general Kasner asymptotics to analytically prove that the Null Energy Condition is sufficient to guarantee the existence of this intermediate critical extremal surface $\mathcal{A}_c$, which is then explicitly verified within an Einstein-scalar gravity model.
Geometrically, this means that as $\tau_0 \to \infty$, the tip of the bulk extremal surface extends deeper into the interior and lies arbitrarily close to this critical surface $\mathcal{A}_c$ located at $z_{c}$. In other words, the local geometry of the critical surface $\mathcal{A}_c$ at $z_c$ completely dictates the late-time behavior of the TEE.

To analyze this limit, we introduce a small positive parameter $\delta \to 0^+$, such that $H_*^4 = H_{*,\min}^4 + \delta = g(z_c) + \delta$. Near $z=z_c$, we can Taylor expand $g(z)$ to leading order: $g(z) \approx g(z_c) - k^2(z-z_c)^2$ (where $k^2 = -g''(z_c)/2 > 0$ since it is a maximum). Consequently, the function $X(z)$ behaves near $z_c$ as:
\begin{equation}
        X(z) \approx z_c^{d-1} \sqrt{\delta + k^2(z-z_c)^2}\ .
\end{equation}
Both the parametric integrals for $\tau_0(H_*)$ in Eq.~\eqref{eq:tau_0&H_*} and the real area density $\text{Re}\mathcal{A}{A_1A_2}(H_*)$ in Eq.~\eqref{eq:A&H_*} contain a $1/X$ factor in their integrands. This $\int \frac{\dd z}{\sqrt{\delta+\dots}}$ structure naturally induces a logarithmic divergence of the form $\sim -\ln \delta$ as $\delta \to 0$:
\begin{equation}\label{eq:tau0AHlim}
    \begin{aligned}
        \tau _0(H_*) & \approx \frac{2z_{c}^{2-2d}}{\e^{\chi \left( z_c \right) /2}H_{*,\min}^{2}k}\ln \delta ^{-1}\ ,\\
        2\text{Re}\mathcal{A}_{A_1A_2}(H_*) & \approx \frac{2z_{c}^{2-2d}}{\e^{\chi \left( z_c \right) /2}k}\ln \delta ^{-1}\ .
    \end{aligned}
\end{equation}
This confirms that both the boundary temporal width $\tau_0$ and the real area diverge $\tau_0 \to \infty$ logarithmically as $H_* \to H_{*,\min}$. By eliminating the divergent $\ln \delta^{-1}$ term between the two asymptotic expressions in Eq.~\eqref{eq:tau0AHlim}, we find a direct, universal linear relationship at late times:
\begin{equation}\label{eq:linearReAtau}
    2\text{Re}\mathcal{A}_{A_1A_2}(\tau _0) \approx H_{*,\min}^{2}\tau_0+\text{finite part}\ , \quad \text{as} \quad \tau_0 \to \infty\ .
\end{equation}
Since the imaginary part of the TEE is independent of $\tau_0$ (Sec.~\ref{subsubsec:imagipart}), the dynamical growth rate of the total TEE is entirely governed by its real part. From Eq.~\eqref{eq:linearReAtau}, we can directly read off the late-time growth rate limit:
\begin{equation}\label{eq:integralbound}
    \lim_{\tau_0 \to \infty}\frac{\dd \mathcal{A}_{A_1 A_2B_2 B_1}(\tau _0)}{\dd \tau_0}= H_{*,\min}^{2}= \sqrt{g(z_c)}\ .
\end{equation}
This result demonstrates that the late-time growth rate approaches the constant $\sqrt{g(z_c)}$, where $z_c$ is the point at which $g(z)$ attains its global maximum. The exact location $z_c$ of $\mathcal{A}_c$ is implicitly determined by the geometric condition $g'(z=z_c)=0$ for any given blackening factor $f(z)$:
\begin{equation}\label{eq:locationz_c}
    \left(\chi'\left(z_c\right) z_c+2 d-2\right) f \left(z_c\right)=f'\left(z_c\right) z_c\ .
\end{equation}
Remarkably, this equation is equivalent to the on-shell condition $\partial_z L_\text{on-shell}|_{z=z_c, z'=0} = 0$, where $L_\text{on-shell}$ is the on-shell value of Lagrangian~\eqref{eq:Lagrangian}.

\subsubsection{Method 2: Hamilton-Jacobi Formalism}

Alternatively, the Hamilton-Jacobi formalism directly extracts the growth rate without relying on explicit asymptotic expansions. We can regard the area functional of the space-like segment $A_1 A_2$ as an on-shell action:
\begin{equation}\label{eq:on-shell}
    \mathcal{S}_{A_1 A_2}(v_{A_1}) \equiv\mathcal{A}_{A_1 A_2}(v_{A_1}) =\int^{v_{A_2}=0}_{v_{A_1}=\tau_0/2} \dd v\,z^{1-d}(v) \sqrt{-f(z) \e^{-\chi (z)}-2\e^{-\chi (z) /2}z'}\ ,
\end{equation}
Using the standard Hamilton-Jacobi formalism, the derivative of the on-shell action with respect to the boundary ``time'' coordinate $v_{A_1}$ yields the conjugate momentum. This momentum corresponds exactly to the conserved quantity $E$ defined previously in Eq.~\eqref{eq:conservedH}: $\partial\mathcal{S}_{A_1 A_2 B_2 B_1}/\partial v_{A_1}= -2E|_{A_1}$. Here $E|_{A_1}$ is a constant along the extremal surface $A_1 A_2$ and $B_1 B_2$. Using the chain rule, the growth rate of the dynamic real part of TEE ($2\mathcal{S}_{A_1A_2}$) is directly related to this conserved quantity evaluated at the boundary:
\begin{equation}\label{eq:growthrate1}
    \frac{\dd \mathcal{A}_{A_1 A_2 B_2 B_1}}{\dd \tau_0} = 2 \frac{\dd \mathcal{S}_{A_1 A_2}}{\dd \tau_0} = -E|_{A_1}\ .
\end{equation}
Since energy $E|_{A_1}$ is  conserved along the entire trajectory, we can conveniently evaluate it at any point on the surface, rather than solely at the boundary $A_1$.
For any finite boundary time width $\tau_0$, the bulk extremal surface continuously penetrates the interior without any turning point\footnote{This represents a crucial geometric distinction from standard space-like  HEE. In two-sided TFD setups, such as the Hartman-Maldacena surface~\cite{Hartman:2013qma, Li:2022cvm}, the space-like extremal surface connects two distinct asymptotic boundaries. Consequently, the symmetry of the background geometry mandates the existence of a turning point $z_M$ within the bulk interior where $z' = 0$. In stark contrast, single-sided TEE configuration anchors  to one boundary and plunges directly towards the singularity.}, meaning $X(z) >0$ in Eq.~\eqref{eq:zprime} is strictly positive and $z' \neq 0$ everywhere along the path.
However, in the late-time limit as $\tau_0 \to \infty$, the tip of the extremal surface asymptotically approaches the critical surface $\mathcal{A}_c$ located $z_c$. Therefore, an arbitrarily large boundary temporal width ($\tau_0 \to \infty$) can only be achieved if the bulk surface tightly hugs the critical surface $\mathcal{A}_c$ for an extended segment.
In this exact late-time limit, the asymptotic tip of the extremal surface approaches $z_c$, which physically forces the auxiliary function $X(z \to z_c) \to 0$ and the derivative $z' \to 0$. Evaluating $E$ by taking the limit $z \to z_c$ and $z' \to 0$ at this late-time tip limit gives:
\begin{equation}
    \begin{aligned}
        E|_{z\to z_c} &=\lim _{z\rightarrow z_c,\,z^\prime \rightarrow 0}\frac{z^{1-d}\left( f(z)\e^{-\chi (z)}+\e^{-\chi (z)/2}z^{\prime} \right)}{\sqrt{-f(z)\e^{-\chi (z)}-2\e^{-\chi (z)/2}z^{\prime}}}\\
        &=-\sqrt{g(z_c)}\ ,
    \end{aligned}
\end{equation}
where $g(z)= -z^{2-2d}f(z)\e^{-\chi(z)}$ is exactly the function defined in Eq.~\eqref{eq:Hstar}.
Substituting this result back into Eq.~\eqref{eq:growthrate1}, we obtain the late-time growth rate of the time-like entanglement entropy:
\begin{equation}\label{eq:upperbound}
    \lim_{\tau_0 \to \infty} \frac{\dd \mathcal{A}_{A_1 A_2B_2 B_1}}{\dd \tau_0} = \sqrt{g(z_c)} \ .
\end{equation}
This confirms that the maximum growth rate is purely determined by the maximum of $g(z)$. Again, the extremum $\left.\frac{\dd}{\dd z}\sqrt{g(z)}\right|_{z=z_c} = 0$ immediately recovers the geometric equation $g'(z_c)=0$, Eq.~\eqref{eq:locationz_c}, for the critical surface $\mathcal{A}_c$.

Therefore, both the asymptotic expansion of the integral~\eqref{eq:integralbound} and the Hamilton-Jacobi formalism~\eqref{eq:upperbound} converge to the exact same conclusion: the late-time linear growth rate of TEE is governed by the local geometry at the critical surface $\mathcal{A}_c$.

Remarkably, the geometric equation $g'(z_c)=0$ determining the location of $\mathcal{A}_c$ is algebraically identical to the one governing the Hartman-Maldacena (HM) surface, which computes the standard spatial holographic entanglement entropy for the two-sided TFD state~\cite{Hartman:2013qma, Li:2022cvm}. Because their late-time growth rates are controlled by the exact same local interior geometric quantity $g(z)=-z^{2-2d}f(z)\e^{-\chi(z)}$ or $\sqrt{g(z_c)}$, we prove in Appendix~\ref{appendix:DEC} that, under the dominant energy condition, $g(z)$ is bounded from above by its Schwarzschild-AdS counterpart at a fixed horizon radius and, for asymptotically Schwarzschild-AdS geometries, at a fixed mass density. Taking the maximum over the interior therefore  establishes the corresponding upper bound for the late-time growth rate of the real part of the TEE.

Inspired by this established bound for the space-like contribution, a compelling physical question naturally arises: does a similar bound exist for the imaginary (time-like) part of the TEE? We will systematically investigate this conjecture through numerical computations in Sec.~\ref{sec:ImagofTEE}.

\subsection{Existence of the Critical Extremal Surface}\label{subsec:existence}

In our preceding analysis, we explicitly assumed the existence of a well-defined limit $H_* \to H_{*,\min}$. Geometrically, this requires an intermediate critical extremal surface $\mathcal{A}_c$ located at $z_c$ inside the black hole interior.
However, this existence is not automatically guaranteed and requires a careful investigation of the asymptotic properties near the spacetime singularity. Mathematically, proving the existence of $\mathcal{A}_c$ is equivalent to demonstrating that the geometric function $g(z) = -z^{2-2d}f(z)\e^{-\chi(z)}$, which is continuous on $(z_h,\infty)$, possesses at least one global maximum within the interior region $z_h < z < \infty$.

At the event horizon, $f(z_h)=0$, which yields $g(z_h)=0$. In the region  inside the horizon ($z_h < z < \infty$), the blackening factor is negative ($f(z) < 0$), ensuring that $g(z) > 0$. Therefore, a sufficient condition for $g(z)$ to reach a positive global maximum at some intermediate point $z_c$ is that it vanishes at the singularity, i.e., $\lim_{z \to \infty} g(z) = 0$. While this condition is trivially satisfied in the pure Schwarzschild-AdS and Reissner-Nordstr\"{o}m backgrounds, establishing its universal validity requires a careful investigation of the general asymptotic properties near the spacetime singularity. We will demonstrate that null energy condition (NEC) is sufficient to supports the existence of the critical extremal surface $\mathcal{A}_c$. This is followed by an explicit verification in an Einstein-scalar gravity model.

\subsubsection{Kasner Asymptotics and Geometric Constraints}\label{subsubsec:generalKasner}

To evaluate the limit $\lim_{z \to \infty} g(z)$, it is instructive to consider the asymptotic spacetime near the space-like singularity. Suppose that near the deep interior singularity ($z \to \infty$), the metric functions exhibit the power-law and logarithmic behaviors, $f(z) \sim -f_\infty z^{2(a+1)}$ and $\chi(z) \sim -2b\ln z+\chi_\infty$ (for some $a,b\in \mathbb{R}$ and $f_\infty>0$). Note that since we restrict our analysis to spacetimes featuring a space-like singularity without an inner horizon, the requirement of geodesic incompleteness and the divergence of the Kretschmann scalar\footnote{The proper time along a radial time-like geodesic scales as $\tau = \int^\infty \dd z\,\sqrt{-g_{zz}} \sim \int^\infty \dd z\,z^{-(a+2)}$. For this proper time to be finite at the singularity $z \to \infty$, we must have $a > -1$. Furthermore, the Kretschmann scalar $K = R_{\mu\nu\rho\sigma}R^{\mu\nu\rho\sigma}$ diverges for $a>-1$ as a power-law $\mathcal{O}\left(z^{4(a+1)}\right)$.} physically constrain the parameter to $a > -1$.  

Under these conditions, we assume that near such a space-like singularity, the leading-order metric effectively takes a Kasner form~\cite{Belinski:1973zz, Kasner:1921zz}:
\begin{equation}\label{eq:Kasner}
    \dd s^{2} = -\dd \tau^{2} + \tau^{2p_{t}} \dd t^{2} + \tau^{2p_{s}} (\dd x^2+\dd \boldsymbol{y}_{d-2}^{2})\ ,
\end{equation}
where $p_t$, $p_s$ are the associated Kasner exponents, and the singularity is located at proper time $\tau \to 0$.  
However, not all spacetimes exhibit such Kasner scalings. In this work, we focus on backgrounds that do display these standard Kasner-like behaviors. For the Einstein–Maxwell–scalar system, Ref.~\cite{Cai:2020wrp} proved that if the scalar potential function $V(\phi)$ is a transcendental function (e.g., hyperbolic functions), then the asymptotic geometry near the singularity does not take the Kasner form.
Moreover, other studies (e.g., Refs.~\cite{Hartnoll:2020fhc, Zhang:2025tsa, Zhao:2025odj, Zhao:2026mkx}) have shown that more exotic dynamics, such as Kasner oscillations, can occur, which do not follow the simple power-law scaling of Eq.~\eqref{eq:Kasner}. We leave the generalization of our bounds to such exotic space-like singularities for future exploration.

We can map $z$-coordinates in Eq.~\eqref{eq:AdSmetric} to this Kasner frame via the proper time transformation $\tau \sim z^{-(a+1)}$ to get the Kasner exponents in terms of metric scaling parameters $a$ and $b$:
\begin{equation}\label{eq:definitionp_sp_t}
    \begin{aligned}
        p_s &= \frac{1}{a+1}\ ,\\
        p_t &= -\frac{a+b}{a+1}\ .
    \end{aligned}
\end{equation}
Note that the physically constrain $a > -1$ is equivalent to $p_s > 0$.

By substituting these near-singularity scaling relations into our target geometric function $g(z)= -z^{2-2d}f(z)\e^{-\chi(z)}$, we can express its scaling entirely in terms of the proper time $\tau$ and the Kasner exponents:
\begin{equation}\label{eq:boundgz}
    g(\tau) \sim  \tau^{2p_t + 2p_s(d-2)}\ .
\end{equation}
Since $\tau \to 0$ at the singularity, achieving $\lim_{\tau \to 0} g(\tau) = 0$  requires the combined exponent to be non-negative:
\begin{equation}\label{eq:ineqptps}
  p_t + (d-2)p_s \geqslant 0\,.
\end{equation}
This is our central inequality discussed in this section.

To determine whether this geometric bound is satisfied, consider the standard physical constraints imposed on the interior matter fields. In the context of AdS/CFT duality, reasonable bulk spacetimes are generally expected to satisfy the null energy condition (NEC), which is essential for preserving fundamental consistencies such as boundary causality and quantum information inequalities~\cite{Freedman:1999gp, Wall:2012uf}.
Inside the event horizon, the causal roles of the coordinates exchange: the radial coordinate $z$ becomes time-like, while the coordinate $t$ becomes space-like. This metric signature flip allows us to construct distinct null vectors in different planes. For our specific purpose of constraining the Kasner exponents, it turns out we do not need the full set of energy conditions. We evaluate the NEC both in the transverse $z-x^i$ plane (spanned by $z$ and $y_{d-2}$) and the $z-t$ plane.
We could introduce the auxiliary null vector\footnote{Inside the event horizon where $f(z) < 0$, the radial coordinate $z$ becomes time-like, while the coordinate $t$ becomes space-like. Consequently, we can construct normalized orthogonal basis vectors: $e_{\hat{t}}^\mu = \frac{1}{\sqrt{g_{tt}}}(\partial_t)^\mu = (z\e^{\chi/2}/\sqrt{-f}, 0, \dots)$, $e_{\hat{z}}^\mu = \frac{1}{\sqrt{-g_{zz}}}(\partial_z)^\mu = (0, z\sqrt{-f}, \dots)$, and $e_{\hat{x}^i}^\mu = \frac{1}{\sqrt{g_{x^ix^i}}}(\partial_{x^i})^\mu = (0, \dots, z, \dots)$. From these, the null frames in the transverse $z-x^i$ planes are defined as $l^\mu_{(z,x^i)}, k^\mu_{(z,x^i)} = \frac{1}{\sqrt{2}}(e_{\hat{z}}^\mu \pm e_{\hat{x}^i}^\mu)$. The radial null vectors in the $z-t$ plane are defined by $l_{(z,t)}^\mu , k^\mu_{(z,t)} = \frac{1}{\sqrt{2}}(e_{\hat{z}}^\mu \pm e_{\hat{t}}^\mu)$. It is straightforward to verify that these satisfy the standard null conditions $l_\mu l^\mu = 0$, $k_\mu k^\mu = 0$, and $l_\mu k^\mu = -1$.} $l^\mu$, which satisfies $l_\mu l^\mu = 0$. From the Einstein equations $G_{\mu\nu} + \Lambda g_{\mu\nu} = 8\pi T_{\mu\nu}$, the NEC statement $T_{\mu\nu} l^\mu_{(z,x^i)} l^\nu_{(z,x^i)} \geqslant 0$ applied to the transverse $z-x^i$ plane yields (where the prime denotes the derivative with respect to $z$):
\begin{equation}
    \e^{\chi/2} z^{d-1} \frac{\dd}{\dd z} \left[ \e^{-\chi/2} z^{1-d} (f' - f\chi') \right] - \frac{d-1}{z} f \chi' \geqslant 0\ .
\end{equation}
Substituting the asymptotic scaling behaviors, we obtain:
\begin{equation}
    b^2 + (3a+2)b - (a+1)(d - 2a - 2) \leqslant 0\ .
\end{equation}
Translating this into the Kasner exponents we have
\begin{equation}\label{eq:NECzx}
    p_t^2 - (p_s + 1)p_t - (d-2)p_s \leqslant 0\ .
\end{equation}
This inequality restricts $p_t$ to the interval $[p_{t,-}, p_{t,+}]$, where the lower root is $p_{t,-} = \frac{1}{2}\left[ p_s+1 - \sqrt{(p_s+1)^2+4(d-2)p_s} \right]$. Using this lower bound, we evaluate our target combination:
\begin{equation}
    \begin{aligned}
        p_t + (d-2)p_s &\geqslant \frac{1}{2} \left[ 1+(2d-3)p_s - \sqrt{(p_s+1)^2+4(d-2)p_s} \right]\\
        &= \frac{1}{2} \left[ 1+(2d-3)p_s - \sqrt{\left[1+(2d-3)p_s\right]^2 - 4(d-1)(d-2)p_s^2} \right]\ .
    \end{aligned}
\end{equation}
For any spatial dimension $d \geqslant 2$, the term $4(d-1)(d-2)p_s^2$ is non-negative, ensuring that the square root is always less than or equal to the leading term $1+(2d-3)p_s$.

\begin{figure}[htbp]
\centering
\begin{subfigure}[t]{0.47\textwidth}
\centering
\includegraphics[width=\linewidth]{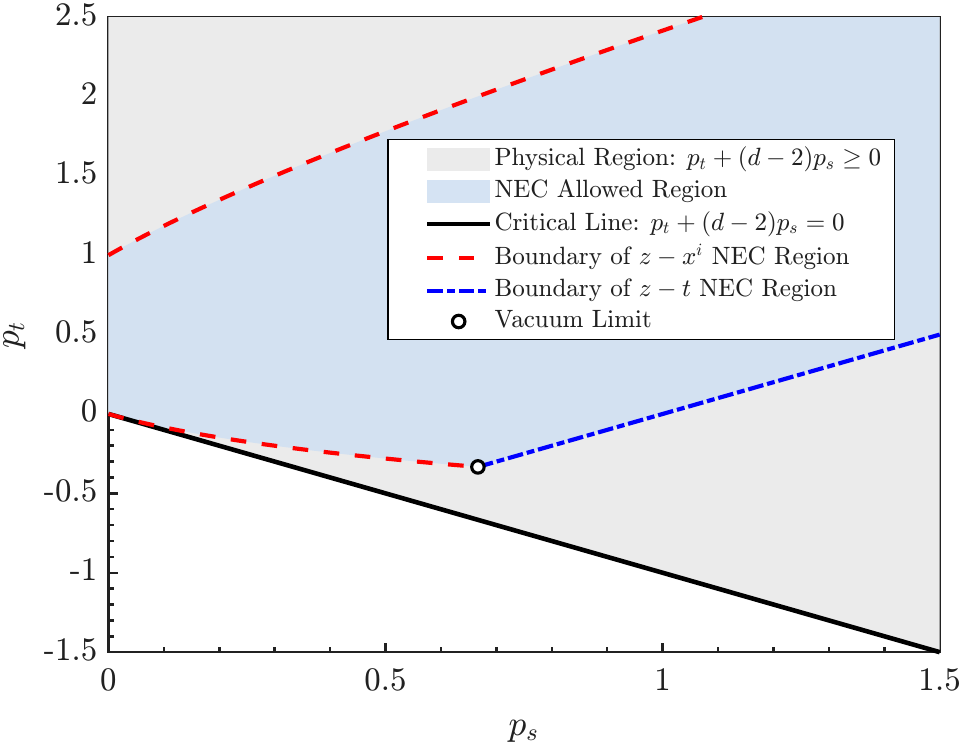}
\caption{$d=3$.}
\end{subfigure}
\qquad
\begin{subfigure}[t]{0.47\textwidth}
\centering
\includegraphics[width=\linewidth]{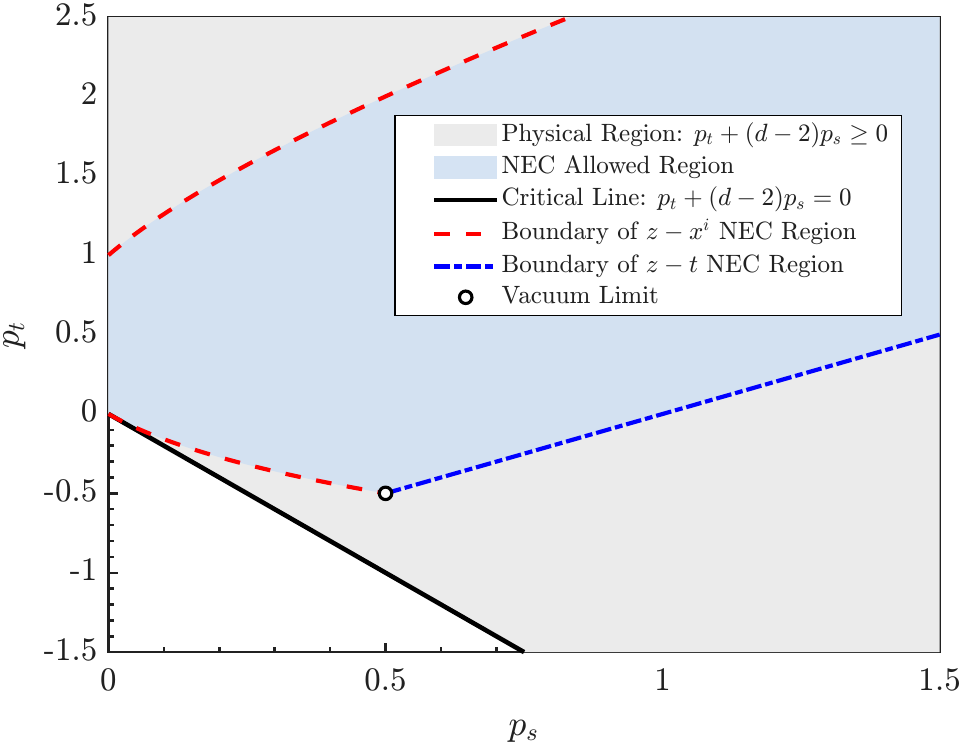}
\caption{$d=4$.}
\end{subfigure}
\caption{Parameter space of the Kasner exponents $p_s$ and $p_t$ near the planar AdS singularity for spatial boundary dimensions (a) $d=3$ and (b) $d=4$. The shaded gray area represents the physically allowed parameter space where the central inequality~\eqref{eq:ineqptps} $p_t + (d-2)p_s \geqslant 0$ holds. The physically allowed parameter space, constrained jointly by the null energy conditions (NEC), is highlighted in light blue. The lower boundary of this valid region is a piece-wise combination of the transverse $z-x^i$ plane NEC (red dashed curve) and the radial $z-t$ plane NEC (blue dash-dotted line). The open circle denotes the vacuum limit $(p_s, p_t) = (2/d, (2-d)/d)$. Notably, the entire NEC-allowed physical region is enveloped by the physically allowed region, maintaining a finite gap above the critical line $p_t + (d-2)p_s = 0$ (black solid line). Thus, the NEC ensures metric combination $g(z) = -z^{2-2d}f(z)\e^{-\chi(z)}$ satisfies $\lim_{z \to \infty} g(z) = 0$. } \label{fig:Kasner1}
\end{figure}

While the transverse NEC in the $z-x^i$ plane is already sufficient to prove our central inequality~\eqref{eq:ineqptps}, for a complete picture of the physically allowed Kasner parameter space, we should also evaluate the NEC in the radial $z-t$ plane. Using the radial null vector $l_{(z,t)}^\mu = \frac{1}{\sqrt{2}}(e_{\hat{z}}^\mu + e_{\hat{t}}^\mu)$, the condition $T_{\mu\nu} l_{(z,t)}^\mu l_{(z,t)}^\nu \geqslant 0$ yields a distinct lower bound:
\begin{equation}
    8\pi T_{\mu\nu} l_{(z,t)}^\mu l_{(z,t)}^\nu = -\frac{d-1}{2} z f(z)\chi^{\prime}(z) \geqslant 0\ .
\end{equation}
Since $f(z) < 0$ inside the horizon, this requires $\chi'(z) \geqslant 0$. Given the asymptotic behavior $\chi(z) \sim -2b\ln z$, this restricts the parameter to $b \leqslant 0$. Translated to the Kasner exponents, we obtain $p_s - p_t = 1 + \frac{b}{a+1}\leqslant 1$ and
\begin{equation}\label{eq:NECzt}
    p_t \geqslant p_s - 1\ .
\end{equation} 
The radial and transverse NEC constraints together define the combined, physically allowed region. Consequently, within the entire physically permitted range ($a > -1$, and thus $p_s >0$), the central inequality~\eqref{eq:ineqptps} holds. This geometrically constrained parameter space is visually summarized in Fig.~\ref{fig:Kasner1}.

Finally, we remark on the theoretical robustness of these geometric bounds against potential quantum gravity corrections. In our preceding analysis, we assumed that classical Einstein gravity remains valid down to the deep interior. However, near the spacelike singularity, quantum gravity effects are expected to become prominent and may lead to violations of classical energy conditions, including the NEC. Fortunately, our central conclusion remains valid as long as quantum gravity corrections preserve the Kasner form of the singularity.
As illustrated in Fig.~\ref{fig:Kasner1}, our target critical line $p_t + (d-2)p_s = 0$ does not strictly coincide with the NEC boundaries. Instead, there exists a finite ``buffer zone'' between the NEC-allowed parameter space (the light blue region) and the critical line. Therefore, a perturbative or mild violation of the NEC near the singularity will only slightly shift the physical boundaries downward, without crossing the critical line. This ``buffer zone'' ensures that the critical surface $\mathcal{A}_c$ still exists, and therefore the late-time TEE growth rate remains well defined, even when classical energy conditions are mildly violated.

Consequently, across all physically permitted ranges of $a > -1$ or $p_s >0$, the null energy condition alone is sufficient to reach the conclusion that $\lim_{\tau \to 0} g(\tau) = 0$. This guarantees the existence of the critical extremal surface $\mathcal{A}_c$. This general proof validates our geometric assumptions for calculating the late-time TEE growth rate. In the next subsection, we will verify that the Einstein-scalar gravity model inherently satisfies this crucial constraint.  

\subsubsection{Explicit Verification in Einstein-Scalar Gravity}

To concretely demonstrate the general result established above, we now explicitly evaluate an Einstein-scalar gravity model. The bulk action is:
\be\label{eq:modelsetup}
\begin{aligned}
 S  & = \frac{1}{16\pi}\int \dd^D x\,\sqrt{-g}\,\left(R-2\Lambda+\L_{\text{m}}\right)\ ,\\
 \L_{\text{m}} &=-\frac{1}{2} \nabla_\mu \phi \nabla^\mu \phi-V(\phi)\ ,
\end{aligned}
\ee
where $R$ is the Ricci scalar, the cosmological constant is $\Lambda= -d(d-1)/2$, and the scalar potential $V(\phi)$ is assumed to be a smooth function of $\phi$. The standard canonical kinetic term $-\frac{1}{2}(\partial\phi)^2$ guarantees that the matter sector respects the NEC everywhere. The corresponding equations of motion are:
\be\label{eq:emtensor}
\begin{aligned}
    &\nabla_\mu \nabla^\mu \phi =\frac{\dd}{\dd \phi}V(\phi)\ ,\\
    &G_{\mu\nu}+ \Lambda g_{\mu\nu}=8\pi T_{\mu\nu}^\text{m}\ ,\\
    &T_{\mu\nu}^\text{m} =\L_{\text{m}}g_{\mu\nu}+\nabla_\mu \phi\nabla_\nu \phi\ ,
\end{aligned}
\ee
where $G_{\mu\nu}$ and $T_{\mu\nu}^\text{m}$ are the Einstein and matter energy-momentum tensors, respectively. Focusing on static planar ($k=0$) black holes, we adopt the standard ansatz $\phi=\phi(z)$, which yields a coupled system of ordinary differential equations as follows:
\be\label{eq:scalareom1}
\begin{aligned}
    \chi ^{\prime}(z) - \frac{z}{d-1}\left(\phi ^{\prime}(z)\right)^2 &= 0\ ,\\
    \frac{V(\phi )}{zf(z)}-\frac{\left( d-1 \right) f^{\prime}\left( z \right)}{f\left( z \right)}+\frac{d\left( d-1 \right)}{z}+\frac{z}{2}\left( \phi^\prime \left( z \right) \right) ^2 &=0\ ,\\
    f(z)\phi ^{\prime \prime}(z)+\left( f^{\prime}(z)+\frac{(1-d)f(z)}{z}-\frac{1}{2}f(z)\chi ^{\prime}(z) \right) \phi ^{\prime}(z)-\frac{1}{z^2}\frac{\dd V(\phi )}{\dd \phi} &=0 \ .
\end{aligned}
\ee

As demonstrated in Ref.~\cite{Cai:2020wrp} by using the  Einstein equations~\eqref{eq:scalareom1}, when the scalar kinetic term dominates the deep interior dynamics ($z \to \infty$), we can safely neglect the potential $V(\phi)$, leading to the exact scaling behaviors:
\be\label{eq:scaling}
\begin{aligned}
    \phi(z)& =\sqrt{d-1}\,\alpha \ln z +\cdots\ ,\\
    \chi (z) &= 2\alpha^2 \ln z+\cdots\ ,\\
    f(z) &= -f_s z^{\alpha^2+d}+\cdots\ .
\end{aligned}
\ee
where $\alpha$, and $f_s$ are constants.

By performing a coordinate transformation from $z$ to the proper time $\tau \sim z^{-(\alpha^2+d)/2}$, one can explicitly verify that the near-singularity metric recovers the standard Kasner form~\eqref{eq:Kasner} with a logarithmic scalar profile $\phi (\tau) =-p_{\phi}\ln \tau$. The corresponding Kasner exponents are fully determined by the parameter $\alpha$:
\be\label{eq:Kasnerexponent}
\begin{aligned}
    p_t &=\frac{2-d+\alpha ^2}{d+\alpha ^2}\ ,\\
    p_s &=\frac{2}{d+\alpha ^2}\ ,\\
    p_{\phi} &=\frac{2\sqrt{d-1}\alpha}{d+\alpha ^2}\ .
\end{aligned}
\ee
Consequently, for any real scalar field ($\alpha^2 > 0$) and $d\geqslant 2$, it is easy to check that $(d-1)p_s+p_t = \frac{d+\alpha^2}{d+\alpha^2} = 1$, which satisfies the NEC requirement for the entire parameter space.

Substituting these exponents into the transverse NEC polynomial derived in Eq.~\eqref{eq:NECzx}, we find:
\begin{equation}
    p_t^2 - (p_s + 1)p_t - (d-2)p_s = -\frac{4(d-1)\alpha^2}{(d+\alpha^2)^2}\ .
\end{equation}
For any physical boundary dimension $d \geqslant 2$ and a standard real scalar field ($\alpha^2 > 0$), this quantity is negative. Thus, the Kasner exponents of this specific Einstein-scalar model automatically fall into the physically allowed regions bounded by the null energy conditions, confirming the requirements of our general proof.

With the scaling relations~\eqref{eq:scaling}, the product behaves as $f(z)\e^{-\chi(z)} \propto -z^{d-\alpha^2}$ as $z \to \infty$. Substituting this back into our target function $g(z)$ yields its ultimate asymptotic form:
\begin{equation}\label{eq:asymofgz}
    \begin{aligned}
        g(z) = -z^{2-2d}f(z)\e^{-\chi(z)} \sim z^{2-d-\alpha^2}\ .
    \end{aligned}
\end{equation}
For $d \geqslant 2$ and a kinetically dominated scalar field ($\alpha^2 > 0$), the exponent is strictly negative ($2-d-\alpha^2 < 0$). Consequently, we conclude that in the deep interior limit, $\lim_{z \to \infty} g(z) = 0$.

This provides a rigorous analytical proof that a critical extremal surface $\mathcal{A}_c$ always exists behind the horizon, validating our assumptions for calculating the late-time TEE growth rate.
While this result has been  proven here for Einstein-scalar gravity models\footnote{For the Einstein-Maxwell-scalar system, Ref.~\cite{Cai:2020wrp} proved that if the potential function $V(\phi)$ is an algebraic function (e.g. a polynomial), rather than a transcendental function (e.g., hyperbolic functions), then the asymptotic geometry near the singularity still takes the Kasner form.}, determining whether such an intermediate critical extremal surface universally exists in more exotic setups (e.g., dimensional reductions of super-gravity theories~\cite{Nunez:2025gxq, Nunez:2025puk}, modify the gravitational theories to include higher-curvature terms~\cite{Zhao:2025zgm, Grandi:2021ajl} or non-relativistic holographic theories~\cite{Afrasiar:2024lsi, Afrasiar:2024ldn, Paul:2024lmd}) remains a compelling open question for future investigation.

\section{Time-like Entanglement and its Growth Rate in Hairy Black Holes}\label{sec:ImagofTEE}

In the previous sections, we established the general analytical framework for the TEE and proved that its late-time growth rate is governed by an interior critical extremal surface. As demonstrated in Sec.~\ref{subsec:asymptau}, the equation governing the real part of this growth is algebraically identical to that of the spatial Hartman-Maldacena surface. Further, we provide a direct analytic proof in Appendix~\ref{appendix:DEC} showing that, as long as the bulk spacetime satisfies the dominant energy condition (DEC), the pure vacuum geometry sets a strict upper bound on this real growth rate. Motivated by this analytical guarantee, we proceed to test the robustness of the real-part upper bound in scenarios that violate the DEC, and independently investigate whether the vacuum geometry minimizes the imaginary part of the TEE.

To systematically investigate and test this conjecture, we will construct numerical solutions for a hairy black hole in an Einstein-scalar gravity model. We specify the scalar potential to be a simple mass term, $V(\phi)=\frac{1}{2}m^2\phi^2$ in holographic bulk action~\eqref{eq:modelsetup}, where $m$ is the mass parameter of the real scalar field $\phi$.
To ensure the spacetime remains asymptotically AdS, the metric functions and the scalar field must satisfy the following boundary conditions near the conformal boundary ($z\to 0$):
\begin{equation}
    \lim_{z\to 0} f(z)=1, \quad \lim_{z\to 0}\chi(z)=0, \quad \lim_{z\to 0}\phi(z)=0\ .
\end{equation}
Near this asymptotic AdS boundary, the scalar field admits a standard expansion:
\begin{equation}\label{eq:phiexpAdS}
    \phi(z) = \phi_- z^{\Delta-} + \phi_+ z^{\Delta+}+\cdots\ ,
\end{equation}
where $\phi_-$ and $\phi_+$ are the non-trivial parameters corresponding to the source and the vacuum expectation value of the dual scalar operator in standard quantization scheme, respectively.
According to the standard holographic dictionary, the conformal dimension $\Delta_\pm$ of dual operators in the CFT$_d$ is determined by the bulk mass via $\Delta_\pm=\frac{1}{2}\left(d\pm\sqrt{d^2+4m^2}\right)$.

In an asymptotic AdS background, the squared mass $m^2$ is permitted to be slightly negative without triggering a dynamical instability, provided it satisfies the well-known Breitenlohner-Freedman bound: $m^2 \geqslant m^2_{\text{BF}} \equiv -d^2/4$. Furthermore, when the mass falls within the specific window $m^2_{\text{BF}} \leqslant m^2 \leqslant m^2_{\text{BF}}+1$, both the $\Delta_+$ and $\Delta_-$ modes are normalizable.
This allows us to study two distinct boundary theories: the standard quantization scheme (where $\phi_-$ is the source) and the alternative quantization scheme (where $\phi_+$ is the source).

In this section, we set the bulk dimension to $D=4$ (dual to a $d=3$ CFT) and fix the scalar mass parameter to $m^2 = -2$. However, because our chosen scalar mass squared is negative, the scalar field locally violates the DEC deep within the black hole interior. Therefore, it provides a useful test of the real-part upper bound and a numerical probe of the conjectural vacuum minimum of imaginary part. Identifying the minimal geometric or energy-condition assumptions behind the observed imaginary-part behavior remains an open question.

To quantify the effects induced by the scalar hair, we normalize our numerical results against the pure vacuum planar Schwarzschild-AdS (SAdS) black hole, characterized by the vacuum blackening factor $f_0(z) = 1 - (z/z_h)^d$, with a subscript ``0'' (e.g., $\sqrt{g_0(z_{c,0})}$ for its maximum real growth rate at its own critical surface $z_{c,0}$, and $\text{Im}\mathcal{A}_0$ for its imaginary part). By evaluating the normalized ratios $\sqrt{g(z_c)} / \sqrt{g_0(z_{c,0})}$ and $\text{Im}\mathcal{A} / \text{Im}\mathcal{A}_0$ under the separate thermodynamic constraints of fixed horizon radius (entropy density), fixed temperature and fixed total mass/energy density, we can systematically determine whether the vacuum geometry establishes bounds for time-like entanglement.

\subsection{Fixed Horizon Radius and Fixed Temperature}

We first evaluate the thermodynamic ensembles where either the horizon radius $z_h$ (equivalent to fixing the entropy density $s\equiv S/\V_{d-2}= z^{1-d}_h/4$ of the dual boundary state) or the Hawking temperature $T= -\frac{f'(z_h)\e^{-\chi(z_h)/2}}{4\pi}$ of boundary thermal state is held constant.

\begin{figure}[htbp]
\centering
\begin{subfigure}[t]{0.36\textwidth}
\centering
\includegraphics[width=\linewidth]{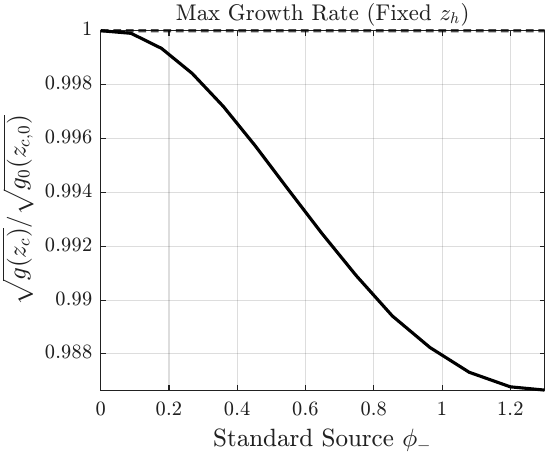}
\caption{}
\end{subfigure}
\qquad
\begin{subfigure}[t]{0.347\textwidth}
\centering
\includegraphics[width=\linewidth]{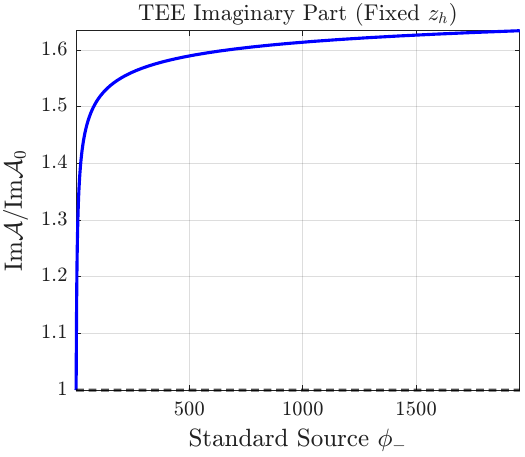}
\caption{}
\end{subfigure}
\caption{Numerical results for the TEE at a fixed horizon radius $z_h$ under the standard quantization scheme. (a) The normalized late-time growth rate $\sqrt{g(z_c)} / \sqrt{g_0(z_{c,0})}$. (b) The normalized imaginary part $\text{Im}\mathcal{A} / \text{Im}\mathcal{A}_0$. The vacuum case ($\phi_- = 0$) acts as an upper bound for (a) and a lower bound for (b) with increasing the scalar source $\phi_-$. }
  \label{fig:imagpartviazh-}
\end{figure}

\begin{figure}[htbp]
\centering
\begin{subfigure}[t]{0.36\textwidth}
\centering
\includegraphics[width=\linewidth]{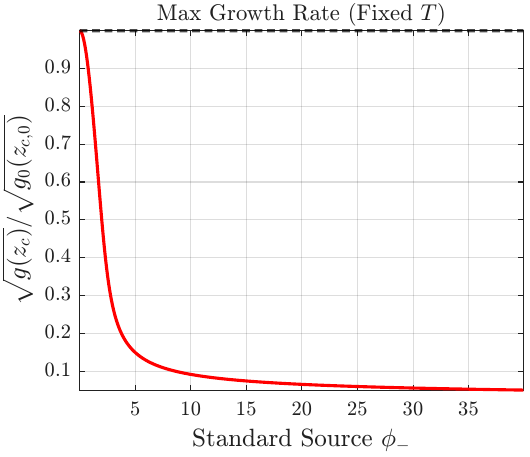}
\caption{\label{fig:GmaxT}}
\end{subfigure}
\qquad
\begin{subfigure}[t]{0.36\textwidth}
\centering
\includegraphics[width=\linewidth]{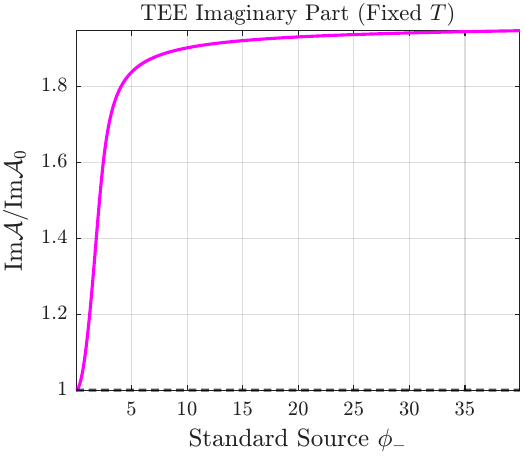}
\caption{\label{fig:ImAT}}
\end{subfigure}
\caption{Numerical results for the TEE at a fixed Hawking temperature $T$ under the standard quantization scheme. (a) The normalized maximum growth rate of the real part ($\sqrt{g(z_c)} / \sqrt{g_0(z_{c,0})}$) is  suppressed by the scalar field, confirming the vacuum upper bound. (b) The normalized imaginary part ($\text{Im}\mathcal{A} / \text{Im}\mathcal{A}_0$) is monotonically enhanced by the scalar field, indicating that the vacuum geometry gives the minimum value at a fixed temperature within this numerical family.}
  \label{fig:imagpartviaT-}
\end{figure}

Under the standard quantization scheme, we explicitly treat the leading-order coefficient $\phi_-$ as the source. By numerically solving the coupled Einstein-scalar equations~\eqref{eq:scalareom1} for different values of $\phi_-$, we generate families of hairy black holes satisfying the respective thermodynamic constraints.
The numerical results for fixed $z_h$ and fixed $T$ are illustrated in Fig.~\ref{fig:imagpartviazh-} and Fig.~\ref{fig:imagpartviaT-}, respectively.

In both ensembles, we observe a robust, universal behavior: as we increase the scalar source $\phi_-$ (injecting more scalar hair into the spacetime), matter backreaction suppresses the normalized late-time growth rate of the real part ($\sqrt{g(z_c)} / \sqrt{g_0(z_{c,0})} \leqslant 1$, see Fig.~\ref{fig:imagpartviazh-}(a) and Fig.~\ref{fig:imagpartviaT-}(a)), while enhancing the imaginary part ($\text{Im}\mathcal{A} / \text{Im}\mathcal{A}_0 \geqslant 1$, see Fig.~\ref{fig:imagpartviazh-}(b) and Fig.~\ref{fig:imagpartviaT-}(b)). The pure vacuum SAdS black hole ($\phi_- = 0$)  acts as the global maximum for the real growth rate and the global minimum for the imaginary part.

\begin{figure}[htbp]
\centering
\begin{subfigure}[t]{0.36\textwidth}
\centering
\includegraphics[width=\linewidth]{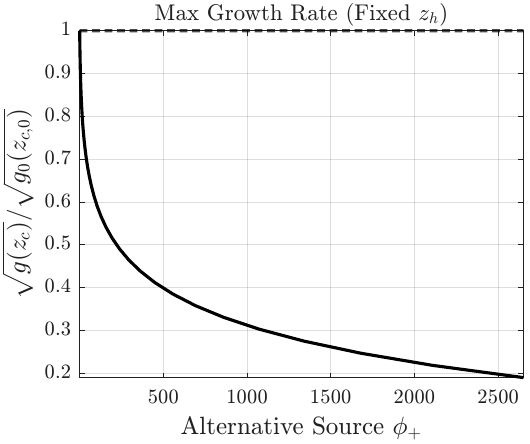}
\caption{\label{fig:Gmaxzh+}}
\end{subfigure}
\qquad
\begin{subfigure}[t]{0.36\textwidth}
\centering
\includegraphics[width=\linewidth]{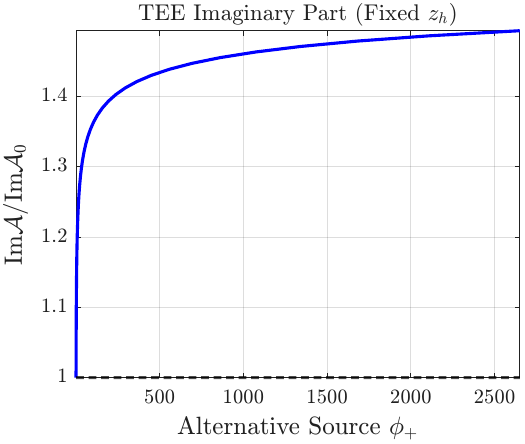}
\caption{\label{fig:ImAzh+}}
\end{subfigure}
\caption{Numerical results for the TEE at a fixed horizon radius $z_h$ under the alternative quantization scheme. (a) The real part's growth rate is suppressed by the scalar field $\phi_+$. (b) The imaginary part is enhanced by the scalar field $\phi_+$. }
  \label{fig:imagpartviazh+}
\end{figure}

\begin{figure}[htbp]
\centering
\begin{subfigure}[t]{0.36\textwidth}
\centering
\includegraphics[width=\linewidth]{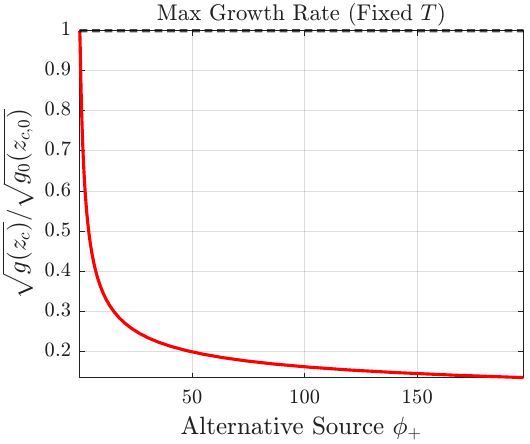}
\caption{\label{fig:GmaxT+}}
\end{subfigure}
\qquad
\begin{subfigure}[t]{0.36\textwidth}
\centering
\includegraphics[width=\linewidth]{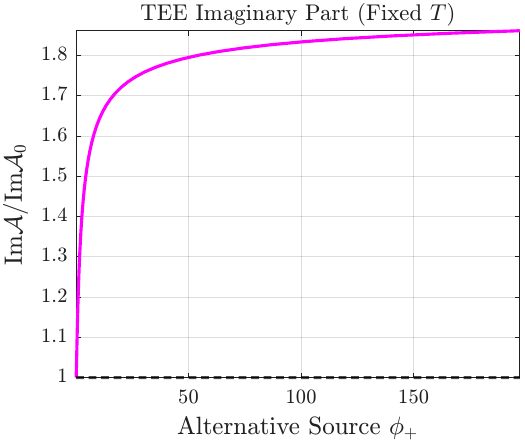}
\caption{\label{fig:ImAT+}}
\end{subfigure}
\caption{Numerical results for the TEE at a fixed Hawking temperature $T$ under the alternative quantization scheme. (a) With temperature fixed, increasing the source parameter $\phi_+$ suppresses the real growth rate. (b) With temperature fixed, increasing $\phi_+$ enhances the imaginary part of the TEE, supporting the numerical vacuum-minimum behavior of the imaginary part.}
  \label{fig:imagpartviaT+}
\end{figure}

This contrast between the real and imaginary parts is physically important. For the real part, the presence of matter fields suppresses the generation of time-like quantum entanglement. For the imaginary part, our numerical findings strongly support our conjecture: just as the vacuum geometry establishes an upper bound for the real part's growth rate under the DEC, it also establishes a universal lower bound for the imaginary part ($\text{Im}\mathcal{A} \geqslant \text{Im}\mathcal{A}_0$).
Remarkably, even though our chosen mass $m^2=-2$ theoretically allows for potential local violations of the DEC deep within the bulk, the lower bound remains robust in all our numerical solutions.

Furthermore, as demonstrated in Fig.~\ref{fig:imagpartviazh+} and Fig.~\ref{fig:imagpartviaT+}, repeating this analysis under the alternative quantization scheme (varying $\phi_+$ as the source) yields the same physical conclusions. Injecting scalar hair into the bulk by increasing $\phi_+$ suppresses the real part's growth rate (Fig.~\ref{fig:imagpartviazh+}(a)) while enhancing the magnitude of the imaginary part (Fig.~\ref{fig:imagpartviazh+}(b)).

Thus, at a fixed entropy density or fixed temperature, the pure vacuum SAdS geometry acts as an upper bound for the real dynamical growth rate and a lower bound for the imaginary part of the time-like entanglement entropy.

\subsection{Fixed Mass/Energy Density}

Previous studies on standard holographic entanglement entropy~\cite{Li:2022cvm} have revealed that while the vacuum geometry provides an upper bound on the growth rate under fixed total mass/energy density in the standard quantization scheme, this bound can be violated in the alternative quantization scheme.
Since the real part of our TEE is governed by the identical local critical equation $g'(z_c)=0$ as the spatial Hartman-Maldacena surface, it is crucial to investigate whether the TEE (both its real and imaginary parts) exhibits a similar strong quantization-scheme dependence.
In this subsection, we will demonstrate that utilizing a recently proposed scheme-independent thermodynamic formulation~\cite{Li:2026dgx} successfully restores this bound for the TEE.

By solving the equations of motion~Eq.~\eqref{eq:scalareom1} near the asymptotic boundary $z\to 0$, the metric functions and the scalar field can be expanded as~\cite{Li:2020spf}:
\begin{equation}\label{eq:asymptexps2}
    \begin{aligned}
        f(z) &=1+\frac{\phi_-^2}{4}z^2+f_3 z^3+\cdots\ , \\
        \chi(z) &=\frac{\phi_-^2}{4}z^2+\frac{2\phi_-\phi_+}{3}z^3+\cdots\ ,\\
        \phi(z) &=\phi_- z+\phi_+ z^2-\frac{\phi_-^3}{8}z^3+\cdots\ .
    \end{aligned}
\end{equation}
To extract the total thermodynamic energy of the system, we must employ the prescription of holographic renormalization. Because there are two independent parameters, $\phi_-$ and $\phi_+$ in Eq.~\eqref{eq:phiexpAdS}, governing the scalar field near the boundary, we can impose different boundary conditions, corresponding to distinct quantization schemes in the dual CFT.
Choosing a different quantization scheme ($\delta\phi_-=0$ or $\delta\phi_+=0$) is equivalent to adding a finite boundary term, i.e. performing a Legendre transform to the bulk action~\cite{Klebanov:1999tb}. This alters the holographic stress-energy tensor, leading to different total energy definitions for the identical bulk geometric background.

For a static spacetime, the energy density $\mathcal{H}$ is given by the $tt$-component of the boundary holographic stress-energy tensor. Beyond the standard and alternative schemes, Ref.~\cite{Li:2026dgx} recently proposed a modified, quantization-scheme-independent definition of energy:
\begin{equation}\label{eq:modifiedenergy}
    \mathcal{H}_\text{modified} = \mathcal{H}_\text{sta/alt}+\frac{d-\Delta}{d}J\left\langle\mathcal{O}\right\rangle\ .
\end{equation}
Here $\mathcal{H}_\text{sta/alt}$ denotes the energy extracted from the $tt$-component of the holographically renormalized stress-energy tensor, $\Delta$ is the scaling dimension of dual operator $\mathcal{O}$ and $\left\langle\mathcal{O}\right\rangle$ is its expectation value, and $J$ represents the source, identified as either $J=\phi_-$ or $J=\phi_+$. They all depend on the chosen quantization scheme.
Evaluating the $tt$-component of the stress-energy tensor with our asymptotic expansion~\eqref{eq:asymptexps2} yields the energy density $\mathcal{H}$ under these three distinct definitions~\cite{Li:2026dgx}:
\begin{equation}\label{eq:threeH}
    \mathcal{H}=\begin{dcases}
    -2f_3+\phi_-\phi_+\ ,  & \text{standard quantization scheme:}\ J=\phi_-\ ; \\
    -2f_3+2\phi_-\phi_+\ , & \text{alternative quantization scheme:}\ J=\phi_+\ ;\\
    -2f_3+\frac{4}{3}\phi_-\phi_+\ ,& \text{quantization-scheme-independent:}\ J=\phi_\pm\ .
 \end{dcases}
\end{equation}

When the scalar field parameters vanish ($\phi_\pm \to 0$), the bulk metric naturally degenerates into the pure vacuum Schwarzschild-AdS black hole. By systematically integrating the equations of motion from the horizon to the boundary, we extract the asymptotic data $\{f_3, \phi_\pm\}$. We then vary the internal parameters while  holding the chosen energy density $\mathcal{H}$ constant to map out the TEE.

\begin{figure}[htbp]
 \begin{center}
   \includegraphics[width=0.6\textwidth]{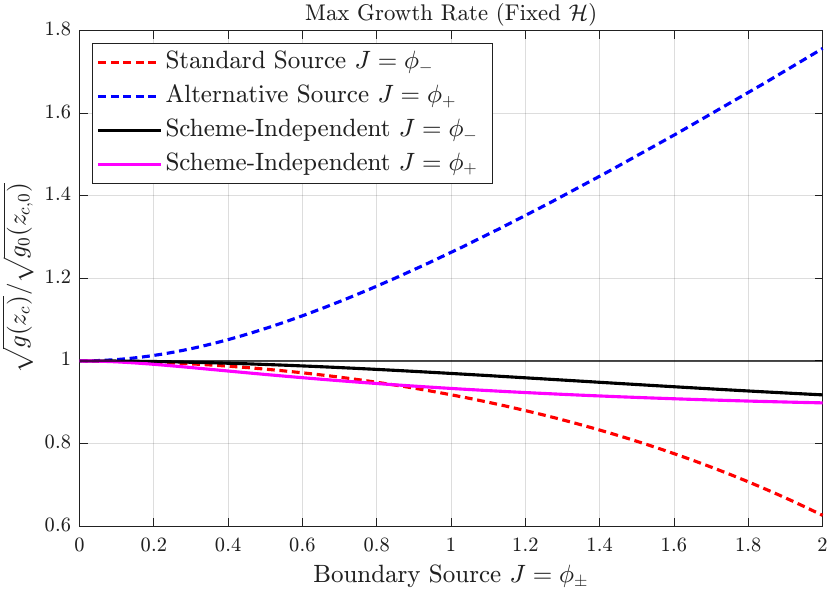}
 \end{center}
\caption{The late-time growth rate of the TEE as a function of the source $J$ under fixed total energy density for different quantization schemes.  The red and blue dashed curves correspond to the standard and alternative quantization schemes (fixing $\mathcal{H}_{\text{sta}}$ and $\mathcal{H}_{\text{alt}}$), respectively. The solid black and purple curves represent the cases where the modified, scheme-independent energy $\mathcal{H}_\text{modified}$ is held constant. } \label{fig:GmaxH}
\end{figure}

\begin{figure}[htbp]
 \begin{center}
   \includegraphics[width=0.6\textwidth]{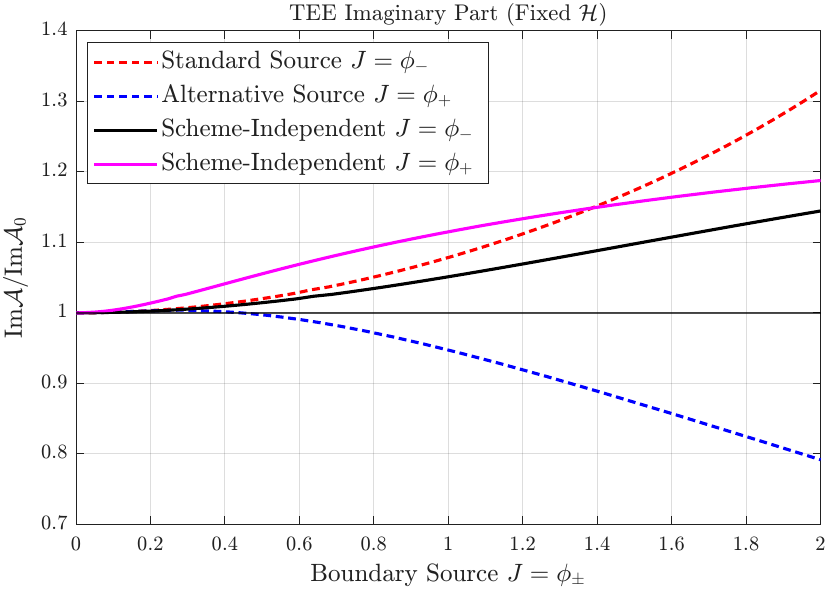}
 \end{center}
\caption{The imaginary part of the TEE as a function of the source $J$ under fixed total energy density for different quantization schemes. The red and blue dashed curves correspond to the standard and alternative quantization schemes, respectively. The solid black and purple curves denote the cases constrained by the scheme-independent energy. \label{fig:ImAH} }
\end{figure}

The numerical results, presented in Fig.~\ref{fig:GmaxH} and Fig.~\ref{fig:ImAH}, provide useful insights. When employing the original energy definitions $\mathcal{H}_\text{sta/alt}$, which explicitly depend on the quantization scheme, we observe a strong dependence of the geometric bounds on the choice of the quantization scheme.
Under the standard quantization scheme (the red dashed curves) for $\mathcal{H}_\text{sta}$ fixed, the bounds hold. The pure vacuum SAdS geometry consistently maximizes the real growth rate and  minimizes the magnitude of the imaginary part.
Under the alternative quantization scheme (the blue dashed curves) for $\mathcal{H}_\text{alt}$ fixed, these bounds are explicitly violated. As the scalar source $\phi_+$ increases, the pure vacuum geometry exhibits a local minimum for the real growth rate and a local maximum for the imaginary part. This implies that the geometric upper and lower bounds for the growth rate and imaginary part fail to universally apply when the total energy is defined via this alternative boundary term.
Strikingly, when the system is constrained by the modified, quantization-scheme-independent energy density (the solid black and purple curves) for $\mathcal{H}_\text{modified}$ fixed, the bounding behaviors are restored for both components. The maximum real growth rate and the minimum imaginary magnitude are once again attained by the pure vacuum geometry.

This confirms that when the total thermodynamic energy is appropriately defined via the scheme-independent formulation as Eq.~\eqref{eq:modifiedenergy}, the  bound properties for time-like entanglement are robustly preserved.
This comprehensive cross-verification solidifies our conclusion: regardless of the quantization scheme or the thermodynamic ensemble chosen, the pure vacuum spacetime configuration universally  maximizes the real dynamical growth rate and universally minimizes the static imaginary part of the Time-like Entanglement Entropy.

\section{Summary and discussion}\label{sec:discussion}

In this paper, we investigated the time evolution of time-like entanglement entropy (TEE) and its late-time growth rate in asymptotically AdS black hole (with space-like singularity and without inner horizon) backgrounds. Analytically, we proved that this late-time growth rate will approach constant and is universally governed by an interior critical extremal surface $\mathcal{A}_c$. This limiting-surface behavior inside the black hole provides an intriguing time-like counterpart to the ``entanglement wedge cosmic censorship'' recently observed in cosmological Kasner backgrounds, where spatial extremal surfaces are similarly driven away from the singularity~\cite{Narayan:2024fcp}. Crucially, by analyzing the asymptotic Kasner geometry near the space-like singularity, we established that the null energy condition (NEC) alone guarantees the existence of this critical surface across all physically permitted parameter spaces, without relying on specific matter models.  Furthermore, we noted that there is a finite gap between the NEC boundaries and the critical line. Because of this gap, the existence of the critical surface, and thus the well-defined TEE growth rate, remains valid even if quantum gravity corrections mildly violate the classical NEC near the singularity.

Numerically, we used a hairy black hole model to test the analytically proven real-part upper bound, specifically extending our examination beyond the DEC assumption and to evaluate the conjectured imaginary-part lower bound. Our results demonstrate that these bounds hold independently of the chosen thermodynamic ensemble: whether the system is constrained at a fixed horizon radius (entropy density) or a fixed Hawking temperature, the pure vacuum Schwarzschild-AdS geometry consistently provides a universal upper bound for the real part and a lower bound for the imaginary part.
Furthermore, we investigated the thermodynamic ensemble with a fixed total energy density, where the holographic energy is defined according to holographic renormalization and inherently depends on the choice of quantum schemes. The validity of geometric bounds in this ensemble is subtle. We found that while they hold under the standard quantization scheme, they are explicitly violated under the alternative scheme. Following the suggestion of Ref.~\cite{Li:2026dgx}, we demonstrated that adopting a scheme-independent energy formulation robustly restores the pure vacuum geometry as the universal extremum for both the real and imaginary components of the TEE.

\begin{figure}[htbp]
 \begin{center}
   \includegraphics[width=0.55\textwidth]{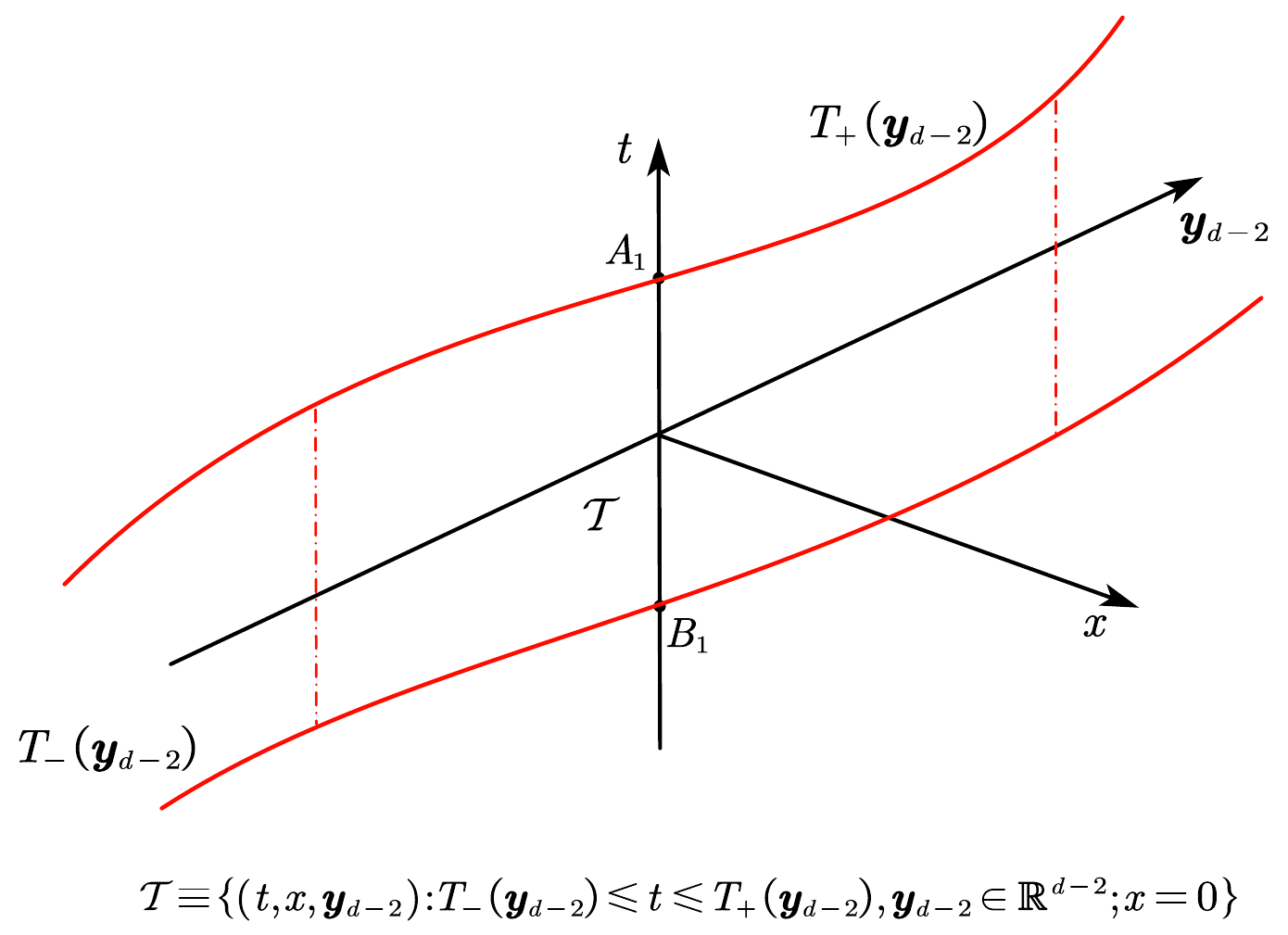}
 \end{center}
\caption{Schematic illustration of an inhomogeneous time-like strip $\mathcal{T}$ on the asymptotic boundary. Unlike the uniform strip studied in the main text, the temporal boundaries $T_\pm(\boldsymbol{y}_{d-2})$ vary along the transverse spatial directions. \label{fig:time-any}}
\end{figure}

Our findings open several compelling avenues for future research. In this work, we focused on a time-like strip with a uniform temporal width $\tau_0$. It is highly instructive to generalize this to an inhomogeneous, arbitrarily shaped boundary time-like strip $\mathcal{T}$, as shown in Fig.~\ref{fig:time-any}:
\begin{equation}
    \mathcal{T} \equiv \left\{ (t,x,\boldsymbol{y}_{d-2}): T_-\left( \boldsymbol{y}_{d-2} \right) \leqslant t \leqslant T_+\left( \boldsymbol{y}_{d-2} \right),\boldsymbol{y}_{d-2}\in \mathbb{R} ^{d-2};x=0 \right\}\ .
\end{equation}
If this inhomogeneous boundary profile evolves linearly with a global time parameter $\tau$, taking the form $T_\pm(\tau;\boldsymbol{y}_{d-2})=\pm\tau+T_\pm(\boldsymbol{y}_{d-2})$, it is natural to conjecture its late-time behavior. As $\tau \to \infty$, We expect that the corresponding bulk extremal surfaces may develop extended portions near a well-defined interior critical surface. Consequently, the TEE might continue to exhibit the universal linear growth properties discussed in this work.
However, extending our framework to such arbitrary configurations introduces two fundamental obstacles. First, it remains to be proven whether the standard Complex-valued Weak Extremal Surface bulk configuration can be consistently constructed for an asymmetric boundary strip.
Second, even if such a generalized CWES configuration is mathematically well-defined, one must then  prove whether a stable critical extremal surface $\mathcal{A}_c$ can actually form in such a less symmetric configuration. Establishing the rigorous existence of these geometric structures, and subsequently determining if the vacuum upper-bound conjecture holds for general time-like entanglements, remains a fascinating open question.

From a gravitational perspective, our numerical setup with $m^2 = -2$ demonstrated that the imaginary lower bound survives even when the DEC is locally violated, but in this case NEC still holds. It is theoretically appealing to push this boundary further by introducing a phantom scalar field. The transformation $\phi \to \i\phi$ reverses the sign of the kinetic term, violating the NEC.
In the deep interior near the singularity, this transformation formally flips the sign of the scaling parameter, $\alpha^2 \to -\alpha^2$, in the asymptotic forms of the metric functions $\chi(z)$ and $f(z)$ Eq.~\eqref{eq:scaling}, and consequently in their combination $g(z)$ defined in Eq.~\eqref{eq:asymofgz}. This algebraic flip effectively acts as an imaginary Kasner exponent $\alpha \to \i\alpha$, which fundamentally alters the near-singularity geometric scaling.
Investigating whether a stable critical extremal surface $\mathcal{A}_c$ can survive such an interior and furthermore, whether the time-like entanglement entropy remains mathematically well-defined, could provide deep insights into the fundamental nature of cosmological singularities and the robustness of TEE. Furthermore, evaluating and comparing how various TEE proposals behave in black holes and Kasner geometries is a significant direction for future research. The presence of bulk matter fields poses a stringent test for the holographic dictionary; for instance, recent work~\cite{Prihadi:2026nua} found that the simple Wick rotation method fails under scalar deformations. This highlights the necessity to systematically test other TEE definitions against non-trivial gravitational backgrounds. 

Finally, from a broader quantum information perspective, the dynamical growth of entanglement is conceptually rooted in fundamental limits of quantum mechanics. For instance, the renowned Lloyd bound~\cite{Lloyd:2000cry} sets a universal upper limit on the growth rate of computational complexity for a given system energy. Moreover, in the context of renormalization group (RG) flows, bounds on the rate of change of time-like entanglement have been elegantly formulated as monotonic holographic c-functions, particularly in Lorentz-violating theories~\cite{Giataganas:2025ize, Giataganas:2025div}. In the broader context of holography, the late-time growth bound of standard spatial entanglement is intimately connected to the evolution of holographic complexity (e.g., via the Complexity=Volume conjecture~\cite{Carmi:2017jqz, Yang:2019alh} and Complexity=Action conjecture~\cite{Brown:2015bva, Brown:2015lvg, Lehner:2016vdi, Cai:2016xho, Yang:2016awy}). Furthermore, as established by our general comparison theorem in Appendix~\ref{appendix:DEC}, the bounding mechanisms for the TEE, spatial entanglement, and certain holographic complexity proposals (such as the CV conjecture) naturally stem from a universal class of interior geometric potentials. Given these geometric and informational connections, it is highly natural to ask if the universal upper bounds on TEE in this work point towards a novel, fundamental notion of upper bounds for ``time-like complexity''~\cite{Alishahiha:2025xml}. Exploring this intricate relationship will be a fascinating direction for future research.

\begin{acknowledgments}
We would like to thank Profs. Haitang Yang, Wu-zhong Guo, and Keun-Young Kim for valuable discussions. We are also grateful to Ze Li, Jaydeep Kumar Basak and Kuntal Pal for helpful discussions. This work is supported by the Natural Science Foundation of China under Grant No. 12375051 and Tianjin University Self-Innovation Fund Extreme Basic Research Project Grant No. 2025XJ22-0014 and 2025XJ21-0007.  
\end{acknowledgments}

\appendix

\section{Alternative Boundary-Evolving Configurations for Time-like Entanglement}\label{appendix:threecase}

\begin{figure}[htbp]
 \begin{center}
   \includegraphics[width=0.9\textwidth]{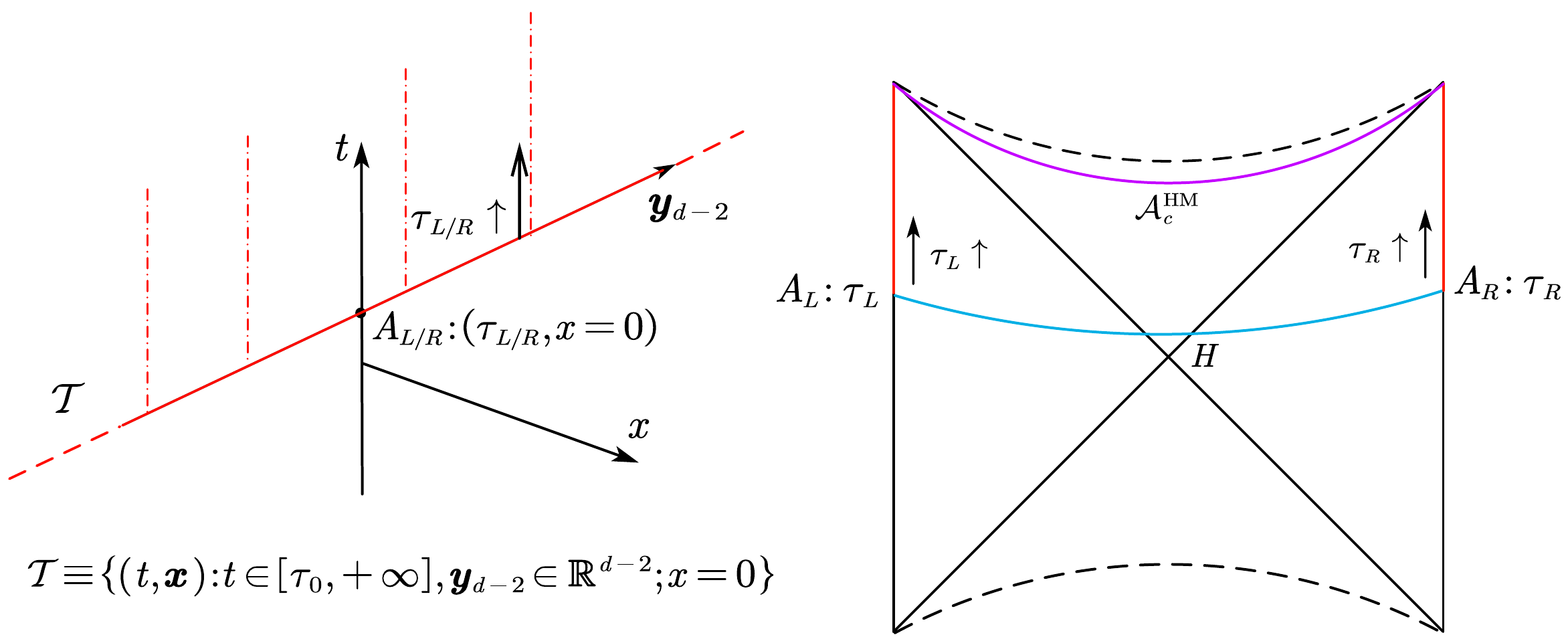}
 \end{center}
\caption{Schematic diagram of the semi-infinite time-like subregions (Case 2) in a two-sided asymptotic AdS black hole geometry. The time-like strips are symmetrically anchored at time coordinate $t=\tau_0$ on both boundaries and extend to future infinity. This configuration evaluates the time-like entanglement entropy in a thermofield double state, where the extremal surfaces are purely space-like and the imaginary part  vanishes.  \label{fig:half-time-space}}
\end{figure}

\begin{figure}[htbp]
 \begin{center}
   \includegraphics[width=0.9\textwidth]{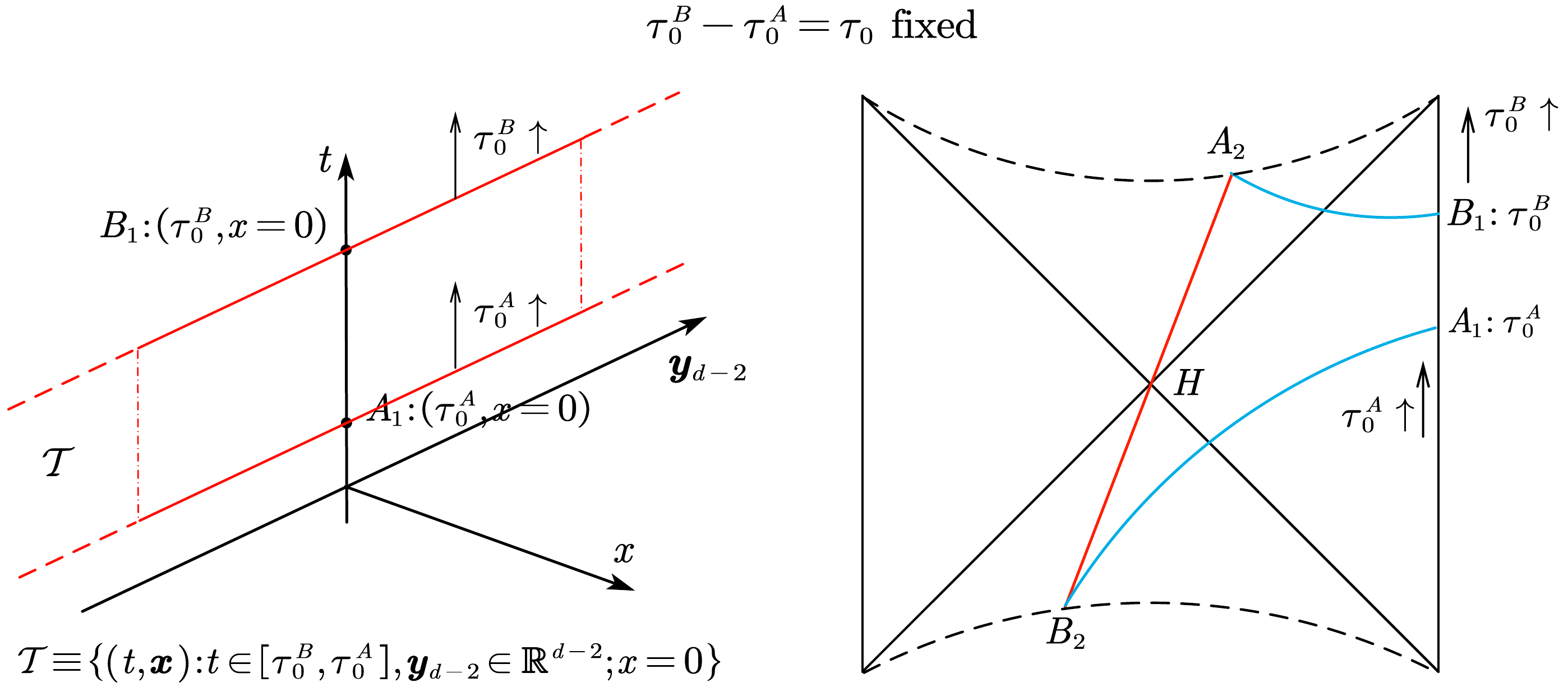}
 \end{center}
\caption{Schematic diagram of the global time translation configuration (Case 3) on a single asymptotic AdS boundary. The time-like strip rigidly translates along the boundary time coordinate while maintaining a  fixed temporal width $\tau_0 = \tau_0^B - \tau_0^A$. \label{fig:asymmetric-time}}
\end{figure}

In this appendix, we discuss three alternative ways of deforming a time-like boundary strip and clarify which of them is relevant for the case discussed in the main text:
\begin{enumerate}
\item \textbf{Symmetric Expansion:} The time-like strip is centered at $t=0$ and extends symmetrically to both the past and future along $t$:
\begin{equation}\label{eq:time-like-strip1}
\mathcal{T}_\text{sym} \equiv \left\{ (t,\boldsymbol{x}): t\in \left[ -\frac{\tau_0}{2},\frac{\tau_0}{2} \right] ,\boldsymbol{y}_{d-2}\in \mathbb{R} ^{d-2};x=0 \right\}\ .
\end{equation}
This is the core configuration investigated in the main text, where TEE captures the accumulation of temporal entanglement, and it grows with $\tau_0$ and approaches the corresponding upper bound at large width $\tau_0\to \infty$.
\item \textbf{Semi-infinite TFD Configuration:} The subregion $\mathcal{T}$ consists of two copies of semi-infinite time-like strips located symmetrically on the two asymptotic AdS boundaries of a two-sided black hole, as illustrated in Fig.~\ref{fig:half-time-space}:
\begin{equation}
    \begin{aligned}
        \mathcal{T}_{\text{TFD}} &\equiv \mathcal{T}_L \cup \mathcal{T}_R \ ,\\
        \mathcal{T}_{L/R} &= \left\{ (t,\boldsymbol{x}): t\in \left[\tau_{L/R}, +\infty\right) ,\boldsymbol{y}_{d-2}\in \mathbb{R} ^{d-2};x=0 \right\}_{L/R}\ .
    \end{aligned}
\end{equation}
This configuration is naturally interpreted in the thermofield-double state. In this case, the relevant extremal surfaces are purely space-like, and the imaginary part of the time-like entanglement entropy vanishes identically~\cite{Li:2022tsv, Anegawa:2024kdj}. Consequently, the real part's growth rate trivially reduces to the well-studied behavior of space-like holographic entanglement entropy~\cite{Hartman:2013qma, Li:2022cvm}.
\item \textbf{Global Time Translation:} A time-like strip of a fixed temporal width $\tau_0$ is located on a single boundary and is globally translated along the boundary time direction:
\begin{equation}\label{eq:time-like-strip2}
\mathcal{T}_\text{trans} \equiv \left\{ (t,\boldsymbol{x}):t\in \left[ \tau_{0}^{A}, \tau_{0}^{B} \right] ,\boldsymbol{y}_{d-2}\in \mathbb{R} ^{d-2};x=0 \right\}\ ,
\end{equation}
where the width $\tau_0^B - \tau_0^A = \tau_0$ remains constant. As shown in Fig.~\ref{fig:asymmetric-time}, the dynamical evolution here is a rigid shift along the time coordinate.
\end{enumerate}

Case 1 is already discussed in the main text. The bound of TEE in case 2 is physically equivalent to  the bound for space-like holographic entanglement entropy. In the following, we restrict our attention to Case 3.

In Case 3, the time-like subregion has a fixed temporal width $\tau_0$ and is rigidly translated along the time direction on the boundary field theory. From the perspective of the dual CFT, due to the continuous time-translation symmetry of the thermal state, we physically expect the time-like entanglement entropy $S_{\mathcal{T}}$ to remain exactly invariant, yielding a zero growth rate.
Here, we provide a rigorous holographic proof from the bulk perspective, demonstrating that the variation of the complex-valued extremal area under this global time translation is identically zero\footnote{The results in Ref.~\cite{Doi:2023zaf} have shown that, in the $d=2$ BTZ case, the geodesic-length expression for TEE is invariant under rigid time translation of a fixed-width time-like interval on the boundary. Here we provide the higher-dimensional proof.} $\dd S_{\mathcal{T}}/\dd\tau^{A/B}_0=0$.

As derived in Sec.~\ref{subsubsec:imagipart}, the imaginary part of the TEE is a geometric constant fixed entirely by the interior bulk geometry, making it independent of the boundary time location. We therefore focus solely on the real part.
For an asymmetrically located strip, the upper space-like segment $A_1 A_2$ is described using the ingoing Eddington coordinate $v$ in Eq.~\eqref{eq:Eddington}, yielding an on-shell action $\mathcal{S}_{A_1 A_2}$ and a conserved conjugate energy $E_v \equiv E_{A_1 A_2}$.
The lower space-like segment $B_1 B_2$ is described using the outgoing Eddington coordinate $u=t+z^*(z)$, where $z^*(z)=\int_0^z{\dd \tilde{z}\,f^{-1}(\tilde{z})\e^{\chi (\tilde{z})/2}}$, yielding an on-shell action $\mathcal{S}_{B_1 B_2}$ and a conjugate energy $E_u \equiv E_{B_1 B_2}$. Because the static metric lacks explicit dependence on $v$ or $u$, both $E_v$ and $E_u$ are conserved along their respective bulk surfaces.

According to Hamilton-Jacobi theory, the variation of the total real area $\mathcal{S}_{A_1 A_2B_1 B_2} = \mathcal{S}_{A_1 A_2} + \mathcal{S}_{B_1 B_2}$ with respect to its endpoints is driven by these conjugate energies. Evaluating the boundary terms at the asymptotic boundary (subscript 1) and the interior singularity junction (subscript 2), we have:
\begin{equation}
    \begin{aligned}
        \delta \mathcal{S}_{A_1 A_2} &= E_v|_{A_2} \delta v_{A_2} - E_v|_{A_1} \delta v_{A_1}\ ,\\
        \delta \mathcal{S}_{B_1 B_2} &= -E_u|_{B_2} \delta u_{B_2} + E_u|_{B_1} \delta u_{B_1}\ .
    \end{aligned}
\end{equation}
Here, the positive and negative signs in the variations originate from the upper and lower limits of the integration respectively, while the variations of the radial coordinates vanish ($\delta z_{A_1} = \delta z_{A_2} = \delta z_{B_1} = \delta z_{B_2} = 0$) because the endpoints are confined to the asymptotic boundary $z=0$ and the singularity $z=\infty$.

We now apply the geometric constraints of rigid translation. On the boundary, the strip's constant width $\tau_0^B - \tau_0^A = \tau_0$ enforces an identical time shift: $\delta v_{A_1} = \delta u_{B_1} \equiv \delta \tau_0^{A/B}$. At the interior singularity, the time-like segment $A_2B_2$ imposes a matching condition $t_{A_2} = t_{B_2}$, which leads to $u_{B_2} - v_{A_2} = 2t_0$, where $t_0=\mathcal{P}\int_0^{\infty}{\dd \tilde{z}\,f^{-1}(\tilde{z})\e^{\chi (\tilde{z})/2}}$ is a constant, with the integral defined in the principal value sense. Thus, the variations at the junction $A_2, B_2$ are also locked together: $\delta v_{A_2} = \delta u_{B_2}$. Grouping the terms yields:
\begin{equation}
    \delta \mathcal{S}_{A_1 A_2B_1 B_2} = (E_v|_{A_2} - E_u|_{B_2}) \delta v_{A_2/B_2} - (E_v|_{A_1} - E_u|_{B_1}) \delta \tau_0^A\ .
\end{equation}

Crucially, for the combined surface to represent a true minimal CWES configuration, the total action must be stationary with respect to the arbitrary junction location at the singularity. This variational principle requires $\partial \mathcal{S}_{A_1 A_2B_1 B_2}/\partial v_{A_2/B_2} = 0$, which immediately forces the difference of the interior energies to vanish: $E_v|_{A_2} - E_u|_{B_2} = 0$. Since the energies are conserved along the surface ($E_v|_{A_1} = E_v|_{A_2}$ and $E_u|_{B_1} = E_u|_{B_2}$), the same difference at the boundary must identically cancel as well:
\begin{equation}\label{eq:cancellation}
    \begin{aligned}
        E_v|_{A_2} - E_u|_{B_2} &= 0\ .\\
        E_v|_{A_1} - E_u|_{B_1} &= 0\ .
    \end{aligned}
\end{equation}
Substituting these cancellations~\eqref{eq:cancellation} back into the total variation equation, we find:
\begin{equation}
 \delta \mathcal{S}_{A_1 A_2 B_2 B_1} = 0 \cdot\delta v_{A_2/B_2} - 0\cdot \delta \tau_0^A = 0  \ .
\end{equation}
Therefore, the derivative with respect to the global boundary time $\tau_0$ exactly vanishes, $\dd S_{\mathcal{T}}/\dd \tau_0^A= 0$ (similarly, $\dd S_{\mathcal{T}}/\dd \tau_0^B= 0$). This confirms that the time-like entanglement entropy remains invariant under global time translations, matching the expectations from the time-translation symmetry of the dual CFT.

\section{A Dominant Energy Condition Comparison Theorem for Interior Growth Potentials}\label{appendix:DEC}

In this appendix, we establish a comparison theorem for a class of geometric potentials that control late-time growth rates behind static AdS black holes. The theorem implies a Schwarzschild-AdS (SAdS) upper bound for any holographic observable whose late-time growth rate is determined by the maximum of one of these potentials. The real part of the TEE, the Hartman-Maldacena surface, and the maximal interior cross-section relevant to the Complexity=Volume proposal provide explicit examples. Throughout this appendix, we continue to set $\ell_\AdS=1$, as in the main text.

Recall the metric in Eq.~\eqref{eq:AdSmetric}, we now temporarily allow $k=0,1$ and 
suppose the spacetime asymptotically approaches a Schwarzschild-AdS black hole:
\begin{equation}
    f(z) = 1+kz^2-\mu z^d + O(z^{d+1})\ , \qquad \chi(z) = O(z^{d+1})\ .
\end{equation} 
For a static transverse metric, the relevant Bondi-Sachs equations reduce to~\cite{Madler:2016xju, Yang:2019alh, Li:2022cvm} 
\begin{align}
    \chi' &= \frac{16\pi z}{d-1}\,T_{zz}\ , \label{eq:appB_chi_equation} \\ 
    -(d-1)z^{d-1} \frac{\dd}{\dd z} \left( z^{-d}f\e^{-\chi/2} \right) &= \e^{-\chi/2} \left[ (d-1)(d-2)k +\frac{d(d-1)}{z^2} -\frac{8\pi}{z^2}(\rho-P) \right]\ . \label{eq:appB_second_equation} 
\end{align} 
Here $T_{zz}$ is the $zz$-component of the matter energy-momentum tensor $T_{\mu\nu}$ in the Bondi–Sachs coordinates. The quantities $\rho=T_{\mu \nu}n^{\mu}n^{\nu}$ and $P=T_{\mu \nu}m^{\mu}m^{\nu}$ represent the energy density and pressure, respectively, measured by an observer with orthogonal time-like and space-like normal vectors $n^{\mu}$ and $m^{\mu}$ of subspace spanned by $\{x^i\}$. The dominant energy condition implies $T_{zz}\geqslant 0$ and  $\rho-P \geqslant 0$. Together with $\chi(0)=0$, the first Eq.~\eqref{eq:appB_chi_equation} gives
\begin{equation}\label{eq:appB_redshift}
   \chi(z)\geqslant 0\ , \qquad 0<\e^{-\chi(z)/2}\leqslant1 \ .
\end{equation}

We now introduce the comparison function
\begin{equation}\label{eq:appB_Y}
    Y(z)=-z^{-d}f(z)\e^{-\chi(z)/2}\ .
\end{equation}
Eq.~\eqref{eq:appB_second_equation} becomes
\begin{equation}\label{eq:appB_Yprime}
    \frac{\dd}{\dd z}Y(z) = \e^{-\chi/2} \left[ \frac{(d-2)k}{z^{d-1}} +\frac{d}{z^{d+1}\ell_{\rm AdS}^2} -\frac{8\pi(\rho-P)}{(d-1)z^{d+1}} \right]\ .
\end{equation}
For the pure Schwarzschild-AdS comparison geometry,
\begin{equation}\label{eq:appB_SAdS}
    f_\text{SAdS}(z) = kz^2+1 -\mu_\text{SAdS}z^d\ , \qquad \chi_\text{SAdS}=0\ ,
\end{equation}
the corresponding function $Y_\text{SAdS}:=-z^{-d}f_\text{SAdS}$ satisfies
\begin{equation}\label{eq:appB_YSprime}
    \frac{\dd}{\dd z}Y_\text{SAdS}(z) = \frac{(d-2)k}{z^{d-1}} +\frac{d}{z^{d+1}}\ .
\end{equation}
Since $d \geqslant 2$ and $k=0,1$, the geometric constraints in Eq.~\eqref{eq:appB_redshift} imply
\begin{equation}\label{eq:appB_derivative_comparison}
    \frac{\dd}{\dd z}Y(z)\leqslant \frac{\dd}{\dd z}Y_\text{SAdS}(z)\ . 
\end{equation}

At fixed horizon radius, we choose $\mu_\text{SAdS}$ such that $f_\text{SAdS}(z_h)=0$. Because $Y(z_h)=Y_\text{SAdS}(z_h)=0$, integration of Eq.~\eqref{eq:appB_derivative_comparison} toward the interior gives
\begin{equation}\label{eq:appB_Y_fixed_horizon}
    Y(z)\leqslant Y_\text{SAdS}(z)\ , \qquad z>z_h.
\end{equation}

At fixed mass density, we set $\mu_\text{SAdS}=\mu$. The asymptotic expansions dictate
\begin{equation}\label{eq:appB_boundary_match} 
    Y(z)-Y_\text{SAdS}(z)=\mathcal{O}(z)\to 0 \qquad z\to0 .  
\end{equation} 
Thus, the difference $D(z)=Y(z)-Y_\text{SAdS}(z)$ obeys $D'(z)\leqslant 0$ and $\lim_{z\to 0}D(z)=0$, which again yields \begin{equation} \label{eq:appB_Y_fixed_mass} 
    Y(z)\leqslant Y_\text{SAdS}(z)\ . 
\end{equation} 
Moreover, if $z$ lies inside the hairy black hole, then $f(z)<0$, meaning $Y(z)>0$. Eq.~\eqref{eq:appB_Y_fixed_mass} therefore implies $Y_\text{SAdS}(z)>0$, which guarantees that this same point also lies inside the SAdS horizon. Thus, at a fixed mass density, we establish the interior domain inclusion $\mathcal I_\text{BH} \subseteq \mathcal I_\text{SAdS}$, where $\mathcal I_\text{BH}$ and $\mathcal I_\text{SAdS}$ denote the respective black hole interior intervals.

We next consider the family of geometric potentials
\begin{equation} \label{eq:appB_general_potential} 
    \mathcal V_{\alpha,\beta}(z) = -Cz^\alpha f(z)\e^{-\beta\chi(z)}\ ,  
\end{equation} 
with $C>0$, $\alpha\in\mathbb R$ and $\beta\geqslant \frac{1}{2}$.
Using Eq.~\eqref{eq:appB_Y}, we can rewrite this potential as
\begin{equation}\label{eq:appB_potential_Y} 
    \mathcal V_{\alpha,\beta}(z) = Cz^{\alpha+d} \e^{-(\beta-\frac{1}{2})\chi(z)}Y(z)\ .  
\end{equation} 
For any $\beta\geqslant\frac{1}{2}$, the exponential factor $\e^{-(\beta-\frac{1}{2})\chi(z)}$ in Eq.~\eqref{eq:appB_potential_Y} is no larger than one. Hence, throughout the black-hole interior, 
\begin{equation} 
    \mathcal{V}_{\alpha,\beta}(z) = -Cz^\alpha f(z)\e^{-\beta\chi(z)} \leqslant \mathcal{V}_{\alpha,\beta}^\text{SAdS}(z) = -Cz^\alpha f_\text{SAdS}(z)\ . \label{eq:appB_main_theorem} 
\end{equation}
This pointwise bound holds for $k=0,1$ under both fixed horizon radius (or entropy density) and fixed mass/energy density constraints.

As discussed in Sec.~\ref{subsec:existence}, given that $\mathcal{V}_{\alpha,\beta}(z=z_h)=0$, the existence of an interior maximum is guaranteed if $\mathcal{V}_{\alpha,\beta}(z\to \infty)$ also vanishes at the space-like singularity, i.e., $\mathcal{V}_{\alpha,\beta}(z\to \infty)=0$. Assuming the same scaling near the space-like singularity as in Sec.~\ref{subsubsec:generalKasner}, $f(z)\sim-f_\infty z^{2(a+1)}$ and $\chi(z)\sim-2b\ln z+\chi_\infty$ (with $f_\infty>0$), we find
\begin{equation}
    \mathcal{V}_{\alpha,\beta}(z) \sim z^{\alpha+2(a+1)+2\beta b}\ .
\end{equation}
Thus the potential vanishes at the singularity if
\begin{equation} 
    \alpha+2(a+1)+2\beta b<0\ . \label{eq:appB_singularity_condition_ab} 
\end{equation} 
In terms of the Kasner exponents defined in Eq.~\eqref{eq:definitionp_sp_t} of Sec.~\ref{subsubsec:generalKasner}, this requirement gracefully translates to: 
\begin{equation} 
    \beta p_t > \frac{\alpha+2\beta}{2}p_s+1-\beta\ . \label{eq:appB_singularity_condition_Kasner} 
\end{equation} 
Whenever Eq.~\eqref{eq:appB_singularity_condition_ab} holds, the potential $\mathcal{V}_{\alpha,\beta}(z)$ is positive in the black hole interior and vanishes at both endpoints. Consequently, it attains a positive maximum at one or more interior critical surfaces $\mathcal{A}_c$, as discussed in Sec.~\ref{subsec:existence}.

We can now state the corresponding result for late-time growth rates. Let $\mathcal{O}$ be a holographic observable whose late-time growth rate admits the representation
\begin{equation} 
    \lim_{t\to\infty}\frac{\dd\mathcal{O}}{\dd t} = \mathcal{K} \sqrt{\underset{z\in \mathcal{I} _{\mathrm{BH}}}{\max}\mathcal{V}_{\alpha,\beta}(z)}\ , \qquad \mathcal{K} > 0\ . \label{eq:unified_observable_rate} 
\end{equation}
The corresponding observable in the SAdS geometry is defined by
\begin{equation} 
    \lim_{t\to\infty}\frac{\dd\mathcal{O}_\text{SAdS}}{\dd t} = \mathcal{K} \sqrt{\underset{z\in \mathcal{I} _{\mathrm{SAdS}}}{\max}\mathcal{V}_{\alpha,\beta}(z)}\ , \qquad \mathcal{K} > 0\ . \label{eq:appB_observable_rate} 
\end{equation}
At fixed horizon radius, the two interior intervals coincide. At fixed mass density, as shown above, we have the interior domain inclusion $\mathcal I_\text{BH}\subseteq\mathcal I_\text{SAdS}$. In both cases, the pointwise bound~\eqref{eq:appB_main_theorem} ensures that the maximum over the hairy black hole interior cannot exceed the maximum over the SAdS interior, therefore implying
\begin{equation} 
    \lim_{t\to\infty}\frac{\dd\mathcal O}{\dd t} \leqslant \lim_{t\to\infty}\frac{\dd\mathcal O_\text{SAdS}}{\dd t} \ . \label{eq:appB_rate_bound} 
\end{equation}
Eq.~\eqref{eq:appB_rate_bound} serves as a powerful conditional comparison theorem: it applies to any observable whose late-time rate adopts the form of Eq.~\eqref{eq:unified_observable_rate}. Determining whether a specific holographic observable belongs to this class requires an independent analysis.

For the real part of the TEE, the relevant potential is
\begin{equation} 
    g(z) = -z^{2-2d}f(z)\e^{-\chi(z)} = \mathcal V_{2-2d,1}(z)\ . \label{eq:appB_TEE_potential} 
\end{equation} 
The late-time growth rate derived in Sec.~\ref{subsec:asymptau} is controlled by the maximum of $g(z)$. Eq.~\eqref{eq:appB_rate_bound} therefore proves that the SAdS geometry provides an upper bound for the late-time growth rate of the real part of the TEE. The same potential governs the Hartman-Maldacena surface. For this specific potential ($\alpha=2-2d$, $\beta=1$), the singularity condition~\eqref{eq:appB_singularity_condition_Kasner} exactly reduces to our central inequality Eq.~\eqref{eq:ineqptps}:
\begin{equation}
p_t+(d-2)p_s>0 .
\end{equation}
Similarly, the maximal interior critical surface relevant to the Complexity=Volume proposal~\cite{Susskind:2014rva, Stanford:2014jda, Yang:2019alh} is governed by the potential $-z^{-2d}f(z)\e^{-\chi(z)} = \mathcal{V}_{\alpha=-2d,\beta=1}(z)$.
 
These examples strongly suggest the existence of a broader universality class of holographic observables whose late-time dynamics are tightly controlled by an interior potential of the form~\eqref{eq:appB_general_potential}. Identifying further members of this class and determining their associated values of $(\alpha,\beta)$ stands as a compelling direction for future work.

\bibliographystyle{JHEP}

\bibliography{TEE-growthrate}

@article{Maldacena:1997re,
    author = "Maldacena, Juan Martin",
    title = "{The Large $N$ limit of superconformal field theories and supergravity}",
    eprint = "hep-th/9711200",
    archivePrefix = "arXiv",
    reportNumber = "HUTP-97-A097, HUTP-98-A097",
    doi = "10.4310/ATMP.1998.v2.n2.a1",
    journal = "Adv. Theor. Math. Phys.",
    volume = "2",
    pages = "231--252",
    year = "1998"
}

@article{Gubser:1998bc,
    author = "Gubser, S. S. and Klebanov, Igor R. and Polyakov, Alexander M.",
    title = "{Gauge theory correlators from noncritical string theory}",
    eprint = "hep-th/9802109",
    archivePrefix = "arXiv",
    reportNumber = "PUPT-1767",
    doi = "10.1016/S0370-2693(98)00377-3",
    journal = "Phys. Lett. B",
    volume = "428",
    pages = "105--114",
    year = "1998"
}

@article{Witten:1998qj,
    author = "Witten, Edward",
    title = "{Anti de Sitter space and holography}",
    eprint = "hep-th/9802150",
    archivePrefix = "arXiv",
    reportNumber = "IASSNS-HEP-98-15",
    doi = "10.4310/ATMP.1998.v2.n2.a2",
    journal = "Adv. Theor. Math. Phys.",
    volume = "2",
    pages = "253--291",
    year = "1998"
}

@article{Maldacena:2001kr,
    author = "Maldacena, Juan Martin",
    title = "{Eternal black holes in anti-de Sitter}",
    eprint = "hep-th/0106112",
    archivePrefix = "arXiv",
    reportNumber = "NSF-ITP-01-59",
    doi = "10.1088/1126-6708/2003/04/021",
    journal = "JHEP",
    volume = "04",
    pages = "021",
    year = "2003"
}

@article{Maldacena:2013xja,
    author = "Maldacena, Juan and Susskind, Leonard",
    title = "{Cool horizons for entangled black holes}",
    eprint = "1306.0533",
    archivePrefix = "arXiv",
    primaryClass = "hep-th",
    doi = "10.1002/prop.201300020",
    journal = "Fortsch. Phys.",
    volume = "61",
    pages = "781--811",
    year = "2013"
}

@article{VanRaamsdonk:2010pw,
    author = "Van Raamsdonk, Mark",
    title = "{Building up spacetime with quantum entanglement}",
    eprint = "1005.3035",
    archivePrefix = "arXiv",
    primaryClass = "hep-th",
    doi = "10.1142/S0218271810018529",
    journal = "Gen. Rel. Grav.",
    volume = "42",
    pages = "2323--2329",
    year = "2010"
}

@article{Ryu:2006bv,
    author = "Ryu, Shinsei and Takayanagi, Tadashi",
    title = "{Holographic derivation of entanglement entropy from AdS/CFT}",
    eprint = "hep-th/0603001",
    archivePrefix = "arXiv",
    reportNumber = "NSF-KITP-06-11",
    doi = "10.1103/PhysRevLett.96.181602",
    journal = "Phys. Rev. Lett.",
    volume = "96",
    pages = "181602",
    year = "2006"
}

@article{Hubeny:2007xt,
    author = "Hubeny, Veronika E. and Rangamani, Mukund and Takayanagi, Tadashi",
    title = "{A Covariant holographic entanglement entropy proposal}",
    eprint = "0705.0016",
    archivePrefix = "arXiv",
    primaryClass = "hep-th",
    reportNumber = "DCPT-07-13, KUNS-2069",
    doi = "10.1088/1126-6708/2007/07/062",
    journal = "JHEP",
    volume = "07",
    pages = "062",
    year = "2007"
}

@article{Calabrese:2005in,
    author = "Calabrese, Pasquale and Cardy, John L.",
    title = "{Evolution of entanglement entropy in one-dimensional systems}",
    eprint = "cond-mat/0503393",
    archivePrefix = "arXiv",
    doi = "10.1088/1742-5468/2005/04/P04010",
    journal = "J. Stat. Mech.",
    volume = "0504",
    pages = "P04010",
    year = "2005"
}

@article{Ho:2015woa,
    author = "Ho, Wen Wei and Abanin, Dmitry A.",
    title = "{Entanglement dynamics in quantum many-body systems}",
    eprint = "1508.03784",
    archivePrefix = "arXiv",
    primaryClass = "cond-mat.stat-mech",
    doi = "10.1103/PhysRevB.95.094302",
    journal = "Phys. Rev. B",
    volume = "95",
    pages = "094302",
    year = "2017"
}

@article{Lloyd:2000cry,
    author = "Lloyd, Seth",
    title = "{Ultimate physical limits to computation}",
    eprint = "quant-ph/9908043",
    archivePrefix = "arXiv",
    doi = "10.1038/35023282",
    journal = "Nature",
    volume = "406",
    pages = "1047--1054",
    year = "2000"
}

@article{Hartman:2013qma,
    author = "Hartman, Thomas and Maldacena, Juan",
    title = "{Time Evolution of Entanglement Entropy from Black Hole Interiors}",
    eprint = "1303.1080",
    archivePrefix = "arXiv",
    primaryClass = "hep-th",
    doi = "10.1007/JHEP05(2013)014",
    journal = "JHEP",
    volume = "05",
    pages = "014",
    year = "2013"
}

@article{Li:2022cvm,
    author = "Li, Ze and Yang, Run-Qiu",
    title = "{Upper bounds of holographic entanglement entropy growth rate for thermofield double states}",
    eprint = "2205.15154",
    archivePrefix = "arXiv",
    primaryClass = "hep-th",
    doi = "10.1007/JHEP10(2022)072",
    journal = "JHEP",
    volume = "10",
    pages = "072",
    year = "2022"
}

@article{Olson:2011bq,
    author = "Olson, S. Jay and Ralph, Timothy C.",
    title = "{Extraction of timelike entanglement from the quantum vacuum}",
    eprint = "1101.2565",
    archivePrefix = "arXiv",
    primaryClass = "quant-ph",
    doi = "10.1103/PhysRevA.85.012306",
    journal = "Phys. Rev. A",
    volume = "85",
    pages = "012306",
    year = "2012"
}

@article{Wang:2018jva,
    author = "Wang, Peng and Wu, Houwen and Yang, Haitang",
    title = "{Fix the dual geometries of $T\bar{T}$ deformed CFT$_2$ and highly excited states of CFT$_2$}",
    eprint = "1811.07758",
    archivePrefix = "arXiv",
    primaryClass = "hep-th",
    reportNumber = "CTP-SCU/2018007",
    doi = "10.1140/epjc/s10052-020-08680-7",
    journal = "Eur. Phys. J. C",
    volume = "80",
    number = "12",
    pages = "1117",
    year = "2020"
}

@article{Doi:2022iyj,
    author = "Doi, Kazuki and Harper, Jonathan and Mollabashi, Ali and Takayanagi, Tadashi and Taki, Yusuke",
    title = "{Pseudoentropy in dS/CFT and Timelike Entanglement Entropy}",
    eprint = "2210.09457",
    archivePrefix = "arXiv",
    primaryClass = "hep-th",
    reportNumber = "YITP-22-121, IMPU22-0052",
    doi = "10.1103/PhysRevLett.130.031601",
    journal = "Phys. Rev. Lett.",
    volume = "130",
    number = "3",
    pages = "031601",
    year = "2023"
}

@article{Doi:2023zaf,
    author = "Doi, Kazuki and Harper, Jonathan and Mollabashi, Ali and Takayanagi, Tadashi and Taki, Yusuke",
    title = "{Timelike entanglement entropy}",
    eprint = "2302.11695",
    archivePrefix = "arXiv",
    primaryClass = "hep-th",
    reportNumber = "YITP-23-22",
    doi = "10.1007/JHEP05(2023)052",
    journal = "JHEP",
    volume = "05",
    pages = "052",
    year = "2023"
}

@article{Li:2022tsv,
    author = "Li, Ze and Xiao, Zi-Qing and Yang, Run-Qiu",
    title = "{On holographic time-like entanglement entropy}",
    eprint = "2211.14883",
    archivePrefix = "arXiv",
    primaryClass = "hep-th",
    doi = "10.1007/JHEP04(2023)004",
    journal = "JHEP",
    volume = "04",
    pages = "004",
    year = "2023"
}

@article{Jiang:2025pen,
    author = "Jiang, Xin and Yang, Haitang",
    title = "{Timelike entanglement entropy revisited}",
    eprint = "2503.19342",
    archivePrefix = "arXiv",
    primaryClass = "hep-th",
    reportNumber = "CTP-SCU/2025004",
    doi = "10.1103/rl9b-373v",
    journal = "Phys. Rev. D",
    volume = "113",
    number = "10",
    pages = "106021",
    year = "2026"
}

@article{Milekhin:2025ycm,
    author        = {Milekhin, Alexey and Adamska, Zofia and Preskill, John},
    title         = {{Observable and computable entanglement in time}},
    eprint        = {2502.12240},
    archiveprefix = {arXiv},
    primaryclass  = {quant-ph},
    month         = {2},
    year          = {2025},
}

@article{Gong:2025pnu,
    author = "Gong, XiangKun and Guo, Wu-zhong and Xu, Jin",
    title = "{Entanglement measures for causally connected subregions and holography}",
    eprint = "2508.05158",
    archivePrefix = "arXiv",
    primaryClass = "hep-th",
    doi = "10.1103/771p-4rkf",
    journal = "Phys. Rev. D",
    volume = "113",
    number = "10",
    pages = "106009",
    year = "2026"
}

@article{Guo:2025ase,
    author = "Guo, Wu-zhong and He, Song and Liu, Tao",
    title = "{Entanglement of General Subregions in Time-Dependent States}",
    eprint = "2512.19955",
    archivePrefix = "arXiv",
    primaryClass = "hep-th",
    month = "12",
    year = "2025"
}

@article{Guo:2024lrr,
    author = "Guo, Wu-zhong and He, Song and Zhang, Yu-Xuan",
    title = "{Relation between time- and spacelike entanglement entropy}",
    eprint = "2402.00268",
    archivePrefix = "arXiv",
    primaryClass = "hep-th",
    doi = "10.1103/gmkp-lrh3",
    journal = "Phys. Rev. D",
    volume = "112",
    number = "8",
    pages = "086020",
    year = "2025"
}

@article{Xu:2024yvf,
    author = "Xu, Jin and Guo, Wu-zhong",
    title = "{Imaginary part of timelike entanglement entropy}",
    eprint = "2410.22684",
    archivePrefix = "arXiv",
    primaryClass = "hep-th",
    doi = "10.1007/JHEP02(2025)094",
    journal = "JHEP",
    volume = "02",
    pages = "094",
    year = "2025"
}

@article{Harper:2025lav,
    author = "Harper, Jonathan and Kawamoto, Taishi and Maeda, Ryota and Nakamura, Nanami and Takayanagi, Tadashi",
    title = "{Non-Hermitian density matrices from timelike entanglement and wormholes}",
    eprint = "2512.13800",
    archivePrefix = "arXiv",
    primaryClass = "hep-th",
    reportNumber = "YITP-25-186",
    doi = "10.1103/j2xw-tcsb",
    journal = "Phys. Rev. D",
    volume = "113",
    number = "12",
    pages = "126017",
    year = "2026"
}

@article{He:2023ubi,
    author = "He, Peng-Zhang and Zhang, Hai-Qing",
    title = "{Holographic timelike entanglement entropy from Rindler method*}",
    eprint = "2307.09803",
    archivePrefix = "arXiv",
    primaryClass = "hep-th",
    doi = "10.1088/1674-1137/ad57a8",
    journal = "Chin. Phys. C",
    volume = "48",
    number = "11",
    pages = "115113",
    year = "2024"
}

@article{Wen:2024yny,
    author = "Wen, Qiang and Xu, Mingshuai and Zhong, Haocheng",
    title = "{Timelike and gravitational anomalous entanglement from the inner horizon}",
    eprint = "2412.21058",
    archivePrefix = "arXiv",
    primaryClass = "hep-th",
    doi = "10.21468/SciPostPhys.18.6.204",
    journal = "SciPost Phys.",
    volume = "18",
    number = "6",
    pages = "204",
    year = "2025"
}

@article{Nunez:2025gxq,
    author = "Nunez, Carlos and Roychowdhury, Dibakar",
    title = "{Timelike entanglement entropy: A top-down approach}",
    eprint = "2505.20388",
    archivePrefix = "arXiv",
    primaryClass = "hep-th",
    doi = "10.1103/vjyt-xc15",
    journal = "Phys. Rev. D",
    volume = "112",
    number = "2",
    pages = "026030",
    year = "2025"
}

@article{Nunez:2025puk,
    author = "Nunez, Carlos and Roychowdhury, Dibakar",
    title = "{Holographic timelike entanglement across dimensions}",
    eprint = "2508.13266",
    archivePrefix = "arXiv",
    primaryClass = "hep-th",
    doi = "10.1007/JHEP11(2025)100",
    journal = "JHEP",
    volume = "11",
    pages = "100",
    year = "2025"
}

@article{Afrasiar:2024lsi,
    author = "Afrasiar, Mir and Basak, Jaydeep Kumar and Giataganas, Dimitrios",
    title = "{Timelike entanglement entropy and phase transitions in non-conformal theories}",
    eprint = "2404.01393",
    archivePrefix = "arXiv",
    primaryClass = "hep-th",
    doi = "10.1007/JHEP07(2024)243",
    journal = "JHEP",
    volume = "07",
    pages = "243",
    year = "2024"
}

@article{Afrasiar:2024ldn,
    author = "Afrasiar, Mir and Basak, Jaydeep Kumar and Giataganas, Dimitrios",
    title = "{Holographic timelike entanglement entropy in non-relativistic theories}",
    eprint = "2411.18514",
    archivePrefix = "arXiv",
    primaryClass = "hep-th",
    doi = "10.1007/JHEP05(2025)205",
    journal = "JHEP",
    volume = "05",
    pages = "205",
    year = "2025"
}

@article{Afrasiar:2025eam,
    author = "Afrasiar, Mir and Basak, Jaydeep Kumar and Kim, Keun-Young",
    title = "{Aspects of holographic timelike entanglement entropy in black hole backgrounds}",
    eprint = "2512.21327",
    archivePrefix = "arXiv",
    primaryClass = "hep-th",
    month = "12",
    year = "2025"
}

@article{Bohra:2025mhb,
    author = "Bohra, Hardik and Sivaramakrishnan, Allic",
    title = "{Composite AdS geodesics for CFT correlators and timelike entanglement entropy}",
    eprint = "2511.22168",
    archivePrefix = "arXiv",
    primaryClass = "hep-th",
    reportNumber = "CALT-TH 2025-037",
    doi = "10.1103/4crg-9y1b",
    journal = "Phys. Rev. D",
    volume = "114",
    number = "2",
    pages = "026015",
    year = "2026"
}

@article{Heller:2024whi,
    author = "Heller, Michal P. and Ori, Fabio and Serantes, Alexandre",
    title = "{Geometric Interpretation of Timelike Entanglement Entropy}",
    eprint = "2408.15752",
    archivePrefix = "arXiv",
    primaryClass = "hep-th",
    doi = "10.1103/PhysRevLett.134.131601",
    journal = "Phys. Rev. Lett.",
    volume = "134",
    number = "13",
    pages = "131601",
    year = "2025"
}

@article{Guo:2025pru,
    author        = {Guo, Wu-zhong and Xu, Jin},
    title         = {{Duality of Ryu-Takayanagi surfaces inside and outside the horizon}},
    eprint        = {2502.16774},
    archiveprefix = {arXiv},
    primaryclass  = {hep-th},
    doi           = {10.1103/xndj-9ftm},
    journal       = {Phys. Rev. D},
    volume        = {112},
    number        = {10},
    pages         = {L101901},
    year          = {2025},
}

@article{Heller:2025kvp,
    author = "Heller, Michal P. and Ori, Fabio and Serantes, Alexandre",
    title = "{Temporal Entanglement from Holographic Entanglement Entropy}",
    eprint = "2507.17847",
    archivePrefix = "arXiv",
    primaryClass = "hep-th",
    doi = "10.1103/qlsv-gp22",
    journal = "Phys. Rev. X",
    volume = "15",
    number = "4",
    pages = "041022",
    year = "2025"
}

@article{Chu:2023zah,
    author        = {Chu, Chong-Sun and Parihar, Himanshu},
    title         = {{Time-like entanglement entropy in AdS/BCFT}},
    eprint        = {2304.10907},
    archiveprefix = {arXiv},
    primaryclass  = {hep-th},
    doi           = {10.1007/JHEP06(2023)173},
    journal       = {JHEP},
    volume        = {06},
    pages         = {173},
    year          = {2023},
}

@article{Chu:2025sjv,
    author        = {Chu, Chong-Sun and Parihar, Himanshu},
    title         = {{Timelike entanglement entropy with gravitational anomalies}},
    eprint        = {2504.19694},
    archiveprefix = {arXiv},
    primaryclass  = {hep-th},
    doi           = {10.1007/JHEP08(2025)038},
    journal       = {JHEP},
    volume        = {08},
    pages         = {038},
    year          = {2025},
}

@article{Li:2026fcr,
    author = "Li, Zi-Hao and Yang, Run-Qiu",
    title = "{Black Hole Interior and Time-like Entanglement Entropy}",
    eprint = "2601.18319",
    archivePrefix = "arXiv",
    primaryClass = "hep-th",
    month = "1",
    year = "2026"
}

@article{Guo:2022sfl,
    author = "Guo, Wu-zhong and He, Song and Zhang, Yu-Xuan",
    title = "{On the real-time evolution of pseudo-entropy in 2d CFTs}",
    eprint = "2206.11818",
    archivePrefix = "arXiv",
    primaryClass = "hep-th",
    doi = "10.1007/JHEP09(2022)094",
    journal = "JHEP",
    volume = "09",
    pages = "094",
    year = "2022"
}

@article{Liu:2022ugc,
    author = "Liu, Bowei and Chen, Hao and Lian, Biao",
    title = "{Entanglement entropy of free fermions in timelike slices}",
    eprint = "2210.03134",
    archivePrefix = "arXiv",
    primaryClass = "cond-mat.stat-mech",
    doi = "10.1103/PhysRevB.110.144306",
    journal = "Phys. Rev. B",
    volume = "110",
    number = "14",
    pages = "144306",
    year = "2024"
}

@article{Omidi:2023env,
    author = "Omidi, Farzad",
    title = "{Pseudo R{\'e}nyi Entanglement Entropies For an Excited State and Its Time Evolution in a 2D CFT}",
    eprint = "2309.04112",
    archivePrefix = "arXiv",
    primaryClass = "hep-th",
    reportNumber = "IPM/P-2023/59",
    month = "9",
    year = "2023"
}

@article{He:2023syy,
    author = "He, Song and Zhang, Yu-Xuan and Zhao, Long and Zhao, Zi-Xuan",
    title = "{Entanglement and pseudo entanglement dynamics versus fusion in CFT}",
    eprint = "2312.02679",
    archivePrefix = "arXiv",
    primaryClass = "hep-th",
    doi = "10.1007/JHEP06(2024)177",
    journal = "JHEP",
    volume = "06",
    pages = "177",
    year = "2024"
}

@article{Mukherjee:2022jac,
    author = "Mukherjee, Jyotirmoy",
    title = "{Pseudo Entropy in U(1) gauge theory}",
    eprint = "2205.08179",
    archivePrefix = "arXiv",
    primaryClass = "hep-th",
    doi = "10.1007/JHEP10(2022)016",
    journal = "JHEP",
    volume = "10",
    pages = "016",
    year = "2022"
}

@article{Katoch:2025bnh,
    author = "Katoch, Gaurav and Sarkar, Debajyoti and Sen, Bhim",
    title = "{Holographic timelike entanglement in AdS$_3$ Vaidya}",
    eprint = "2504.14313",
    archivePrefix = "arXiv",
    primaryClass = "hep-th",
    doi = "10.1103/lmsy-vs86",
    journal = "Phys. Rev. D",
    volume = "112",
    number = "4",
    pages = "046026",
    year = "2025"
}

@article{Katoch:2026dzs,
    author = "Katoch, Gaurav and Sarkar, Debajyoti and Sen, Bhim",
    title = "{Entanglement inequalities for timelike intervals within dynamical holography}",
    eprint = "2604.11158",
    archivePrefix = "arXiv",
    primaryClass = "hep-th",
    month = "4",
    year = "2026"
}

@article{An:2022lvo,
    author = "An, Yu-Sen and Li, Li and Yang, Fu-Guo and Yang, Run-Qiu",
    title = "{Interior structure and complexity growth rate of holographic superconductor from M-theory}",
    eprint = "2205.02442",
    archivePrefix = "arXiv",
    primaryClass = "hep-th",
    doi = "10.1007/JHEP08(2022)133",
    journal = "JHEP",
    volume = "08",
    pages = "133",
    year = "2022"
}

@article{Auzzi:2022bfd,
    author = "Auzzi, Roberto and Bolognesi, Stefano and Rabinovici, Eliezer and Schaposnik Massolo, Fidel I. and Tallarita, Gianni",
    title = "{On the time dependence of holographic complexity for charged AdS black holes with scalar hair}",
    eprint = "2205.03365",
    archivePrefix = "arXiv",
    primaryClass = "hep-th",
    doi = "10.1007/JHEP08(2022)235",
    journal = "JHEP",
    volume = "08",
    pages = "235",
    year = "2022"
}

@article{Belinski:1973zz,
    author = "Belinski, V. A. and Khalatnikov, I. M.",
    title = "{Effect of Scalar and Vector Fields on the Nature of the Cosmological Singularity}",
    journal = "Sov. Phys. JETP",
    volume = "36",
    pages = "591",
    year = "1973"
}

@article{Kasner:1921zz,
    author = "Kasner, Edward",
    title = "{Geometrical theorems on Einstein's cosmological equations}",
    doi = "10.2307/2370192",
    journal = "Am. J. Math.",
    volume = "43",
    pages = "217--221",
    year = "1921"
}

@article{Cai:2020wrp,
    author = "Cai, Rong-Gen and Li, Li and Yang, Run-Qiu",
    title = "{No Inner-Horizon Theorem for Black Holes with Charged Scalar Hairs}",
    eprint = "2009.05520",
    archivePrefix = "arXiv",
    primaryClass = "gr-qc",
    doi = "10.1007/JHEP03(2021)263",
    journal = "JHEP",
    volume = "03",
    pages = "263",
    year = "2021"
}

@article{Li:2020spf,
    author = "Li, Li",
    title = "{On Thermodynamics of AdS Black Holes with Scalar Hair}",
    eprint = "2008.05597",
    archivePrefix = "arXiv",
    primaryClass = "gr-qc",
    doi = "10.1016/j.physletb.2021.136123",
    journal = "Phys. Lett. B",
    volume = "815",
    pages = "136123",
    year = "2021"
}

@article{Freedman:1999gp,
    author = "Freedman, D. Z. and Gubser, S. S. and Pilch, K. and Warner, N. P.",
    title = "{Renormalization group flows from holography supersymmetry and a c theorem}",
    eprint = "hep-th/9904017",
    archivePrefix = "arXiv",
    reportNumber = "CERN-TH-99-86, HUTP-99-A015, USC-99-1, MIT-CTP-2846",
    doi = "10.4310/ATMP.1999.v3.n2.a7",
    journal = "Adv. Theor. Math. Phys.",
    volume = "3",
    pages = "363--417",
    year = "1999"
}

@article{Myers:2010xs,
    author = "Myers, Robert C. and Sinha, Aninda",
    title = "{Seeing a c-theorem with holography}",
    eprint = "1006.1263",
    archivePrefix = "arXiv",
    primaryClass = "hep-th",
    doi = "10.1103/PhysRevD.82.046006",
    journal = "Phys. Rev. D",
    volume = "82",
    pages = "046006",
    year = "2010"
}

@article{Headrick:2007km,
    author = "Headrick, Matthew and Takayanagi, Tadashi",
    title = "{A Holographic proof of the strong subadditivity of entanglement entropy}",
    eprint = "0704.3719",
    archivePrefix = "arXiv",
    primaryClass = "hep-th",
    reportNumber = "SU-ITP-07-08, KUNS-2069, SU-ITP-07/08, KUNS-2069",
    doi = "10.1103/PhysRevD.76.106013",
    journal = "Phys. Rev. D",
    volume = "76",
    pages = "106013",
    year = "2007"
}

@article{Wall:2012uf,
    author = "Wall, Aron C.",
    title = "{Maximin Surfaces, and the Strong Subadditivity of the Covariant Holographic Entanglement Entropy}",
    eprint = "1211.3494",
    archivePrefix = "arXiv",
    primaryClass = "hep-th",
    doi = "10.1088/0264-9381/31/22/225007",
    journal = "Class. Quant. Grav.",
    volume = "31",
    number = "22",
    pages = "225007",
    year = "2014"
}

@article{Klebanov:1999tb,
    author = "Klebanov, Igor R. and Witten, Edward",
    title = "{AdS / CFT correspondence and symmetry breaking}",
    eprint = "hep-th/9905104",
    archivePrefix = "arXiv",
    reportNumber = "PUPT-1863, IASSNS-HEP-99-49",
    doi = "10.1016/S0550-3213(99)00387-9",
    journal = "Nucl. Phys. B",
    volume = "556",
    pages = "89--114",
    year = "1999"
}

@article{Madler:2016xju,
    author = {M{\"a}dler, Thomas and Winicour, Jeffrey},
    title = "{Bondi-Sachs Formalism}",
    eprint = "1609.01731",
    archivePrefix = "arXiv",
    primaryClass = "gr-qc",
    doi = "10.4249/scholarpedia.33528",
    journal = "Scholarpedia",
    volume = "11",
    pages = "33528",
    year = "2016"
}

@article{Carmi:2017jqz,
    author = "Carmi, Dean and Chapman, Shira and Marrochio, Hugo and Myers, Robert C. and Sugishita, Sotaro",
    title = "{On the Time Dependence of Holographic Complexity}",
    eprint = "1709.10184",
    archivePrefix = "arXiv",
    primaryClass = "hep-th",
    doi = "10.1007/JHEP11(2017)188",
    journal = "JHEP",
    volume = "11",
    pages = "188",
    year = "2017"
}

@article{Yang:2019alh,
    author = "Yang, Run-Qiu",
    title = "{Upper bound on cross sections inside black holes and complexity growth rate}",
    eprint = "1911.12561",
    archivePrefix = "arXiv",
    primaryClass = "hep-th",
    doi = "10.1103/PhysRevD.102.106001",
    journal = "Phys. Rev. D",
    volume = "102",
    number = "10",
    pages = "106001",
    year = "2020"
}

@article{Li:2026dgx,
    author = "Li, Ze and Liu, Hai-Shan and Xiao, Zi-Qing and Yang, Run-Qiu",
    title = "{Quantization-scheme-Independent Energy and Its Implications for Holographic Bounds}",
    eprint = "2601.07607",
    archivePrefix = "arXiv",
    primaryClass = "hep-th",
    month = "1",
    year = "2026"
}

@article{Brown:2015bva,
    author = "Brown, Adam R. and Roberts, Daniel A. and Susskind, Leonard and Swingle, Brian and Zhao, Ying",
    title = "{Holographic Complexity Equals Bulk Action?}",
    eprint = "1509.07876",
    archivePrefix = "arXiv",
    primaryClass = "hep-th",
    doi = "10.1103/PhysRevLett.116.191301",
    journal = "Phys. Rev. Lett.",
    volume = "116",
    number = "19",
    pages = "191301",
    year = "2016"
}

@article{Brown:2015lvg,
    author = "Brown, Adam R. and Roberts, Daniel A. and Susskind, Leonard and Swingle, Brian and Zhao, Ying",
    title = "{Complexity, action, and black holes}",
    eprint = "1512.04993",
    archivePrefix = "arXiv",
    primaryClass = "hep-th",
    doi = "10.1103/PhysRevD.93.086006",
    journal = "Phys. Rev. D",
    volume = "93",
    number = "8",
    pages = "086006",
    year = "2016"
}

@article{Lehner:2016vdi,
    author = "Lehner, Luis and Myers, Robert C. and Poisson, Eric and Sorkin, Rafael D.",
    title = "{Gravitational action with null boundaries}",
    eprint = "1609.00207",
    archivePrefix = "arXiv",
    primaryClass = "hep-th",
    doi = "10.1103/PhysRevD.94.084046",
    journal = "Phys. Rev. D",
    volume = "94",
    number = "8",
    pages = "084046",
    year = "2016"
}

@article{Cai:2016xho,
    author = "Cai, Rong-Gen and Ruan, Shan-Ming and Wang, Shao-Jiang and Yang, Run-Qiu and Peng, Rong-Hui",
    title = "{Action growth for AdS black holes}",
    eprint = "1606.08307",
    archivePrefix = "arXiv",
    primaryClass = "gr-qc",
    doi = "10.1007/JHEP09(2016)161",
    journal = "JHEP",
    volume = "09",
    pages = "161",
    year = "2016"
}

@article{Yang:2016awy,
    author = "Yang, Run-Qiu",
    title = "{Strong energy condition and complexity growth bound in holography}",
    eprint = "1610.05090",
    archivePrefix = "arXiv",
    primaryClass = "gr-qc",
    doi = "10.1103/PhysRevD.95.086017",
    journal = "Phys. Rev. D",
    volume = "95",
    number = "8",
    pages = "086017",
    year = "2017"
}

@article{Alishahiha:2025xml,
    author = "Alishahiha, Mohsen",
    title = "{Timelike Holographic Complexity}",
    eprint = "2510.25700",
    archivePrefix = "arXiv",
    primaryClass = "hep-th",
    month = "10",
    year = "2025"
}

@article{Zhao:2025zgm,
    author = "Zhao, Zi-Xuan and Zhao, Long and He, Song",
    title = "{Timelike entanglement entropy in higher curvature gravity}",
    eprint = "2509.04181",
    archivePrefix = "arXiv",
    primaryClass = "hep-th",
    doi = "10.1007/JHEP12(2025)156",
    journal = "JHEP",
    volume = "12",
    pages = "156",
    year = "2025"
}

@article{Grandi:2021ajl,
    author = "Grandi, Nicol{\'a}s and Salazar Landea, Ignacio",
    title = "{Diving inside a hairy black hole}",
    eprint = "2102.02707",
    archivePrefix = "arXiv",
    primaryClass = "gr-qc",
    doi = "10.1007/JHEP05(2021)152",
    journal = "JHEP",
    volume = "05",
    pages = "152",
    year = "2021"
}

@article{Anegawa:2024kdj,
    author = "Anegawa, Takanori and Tamaoka, Kotaro",
    title = "{Black hole singularity and timelike entanglement}",
    eprint = "2406.10968",
    archivePrefix = "arXiv",
    primaryClass = "hep-th",
    doi = "10.1007/JHEP10(2024)182",
    journal = "JHEP",
    volume = "10",
    pages = "182",
    year = "2024"
}

@article{Hartnoll:2020fhc,
    author = "Hartnoll, Sean A. and Horowitz, Gary T. and Kruthoff, Jorrit and Santos, Jorge E.",
    title = "{Diving into a holographic superconductor}",
    eprint = "2008.12786",
    archivePrefix = "arXiv",
    primaryClass = "hep-th",
    doi = "10.21468/SciPostPhys.10.1.009",
    journal = "SciPost Phys.",
    volume = "10",
    number = "1",
    pages = "009",
    year = "2021"
}

@article{Zhao:2026mkx,
    author = "Zhao, Zi-Qiang and Nie, Zhang-Yu and Wei, Shao-Wen and Zhang, Jing-Fei and Zhang, Xin",
    title = "{Interior geometry of black holes as a probe of first-order phase transition}",
    eprint = "2604.01818",
    archivePrefix = "arXiv",
    primaryClass = "gr-qc",
    month = "4",
    year = "2026"
}

@article{Zhao:2025odj,
    author = "Zhao, Zi-Qiang and Nie, Zhang-Yu and Zhang, Xing-Kun and An, Yu-Sen and Zhang, Jing-Fei and Zhang, Xin",
    title = "{Interior structure of black holes with nonlinear terms}",
    eprint = "2512.24893",
    archivePrefix = "arXiv",
    primaryClass = "gr-qc",
    doi = "10.1140/epjc/s10052-026-15625-z",
    journal = "Eur. Phys. J. C",
    volume = "86",
    number = "5",
    pages = "447",
    year = "2026"
}

@article{Zhang:2025tsa,
    author = "Zhang, Xing-Kun and Zhao, Xin and Nie, Zhang-Yu and Hu, Ya-Peng and An, Yu-Sen",
    title = "{Interior structure of the holographic s+p superconductor and chaotic-stable transition near the black hole singularity}",
    eprint = "2506.19419",
    archivePrefix = "arXiv",
    primaryClass = "hep-th",
    doi = "10.1016/j.physletb.2025.140110",
    journal = "Phys. Lett. B",
    volume = "872",
    pages = "140110",
    year = "2026"
}

@article{Prihadi:2026nua,
    author = "Prihadi, Hadyan Luthfan and Al-Faritsi, Muhammad Alifaldi Ramadhan and Firdaus, Rafi Rizqy and Khairunnisa, Fitria and Sarwono, Yanoar Pribadi and Zen, Freddy Permana",
    title = "{Holographic timelike entanglement and subregion complexity with scalar hair}",
    eprint = "2601.18310",
    archivePrefix = "arXiv",
    primaryClass = "hep-th",
    doi = "10.1007/JHEP04(2026)174",
    journal = "JHEP",
    volume = "04",
    pages = "174",
    year = "2026"
}

@article{Paul:2024lmd,
    author = "Paul, Souvik and Roy Chowdhury, Anirban and Saha, Ashis and Gangopadhyay, Sunandan",
    title = "{Information theoretic measures for Lifshitz system}",
    eprint = "2408.03670",
    archivePrefix = "arXiv",
    primaryClass = "hep-th",
    doi = "10.1007/JHEP10(2024)033",
    journal = "JHEP",
    volume = "10",
    pages = "033",
    year = "2024"
}

@article{Paul:2025gpk,
    author = "Paul, Souvik and Guin, Gopinath and Gangopadhyay, Sunandan",
    title = "{Holographic entanglement entropy and complexity for the cosmological braneworld model}",
    eprint = "2505.11553",
    archivePrefix = "arXiv",
    primaryClass = "hep-th",
    doi = "10.1007/JHEP08(2025)164",
    journal = "JHEP",
    volume = "08",
    pages = "164",
    year = "2025"
}

@article{Narayan:2022afv,
    author = "Narayan, K.",
    title = "{de Sitter space, extremal surfaces, and time entanglement}",
    eprint = "2210.12963",
    archivePrefix = "arXiv",
    primaryClass = "hep-th",
    doi = "10.1103/PhysRevD.107.126004",
    journal = "Phys. Rev. D",
    volume = "107",
    number = "12",
    pages = "126004",
    year = "2023"
}

@article{Narayan:2023ebn,
    author = "Narayan, K. and Saini, Hitesh K.",
    title = "{Notes on time entanglement and pseudo-entropy}",
    eprint = "2303.01307",
    archivePrefix = "arXiv",
    primaryClass = "hep-th",
    doi = "10.1140/epjc/s10052-024-12855-x",
    journal = "Eur. Phys. J. C",
    volume = "84",
    number = "5",
    pages = "499",
    year = "2024"
}

@article{Narayan:2023zen,
    author = "Narayan, K.",
    title = "{Further remarks on de Sitter space, extremal surfaces, and time entanglement}",
    eprint = "2310.00320",
    archivePrefix = "arXiv",
    primaryClass = "hep-th",
    doi = "10.1103/PhysRevD.109.086009",
    journal = "Phys. Rev. D",
    volume = "109",
    number = "8",
    pages = "086009",
    year = "2024"
}

@article{Narayan:2024fcp,
    author = "Narayan, K. and Saini, Hitesh K. and Yadav, Gopal",
    title = "{Cosmological singularities, holographic complexity and entanglement}",
    eprint = "2404.00761",
    archivePrefix = "arXiv",
    primaryClass = "hep-th",
    doi = "10.1007/JHEP07(2024)125",
    journal = "JHEP",
    volume = "07",
    pages = "125",
    year = "2024"
}

@article{Takayanagi:2025ula,
    author = "Takayanagi, Tadashi",
    title = "{Essay: Emergent Holographic Spacetime from Quantum Information}",
    eprint = "2506.06595",
    archivePrefix = "arXiv",
    primaryClass = "hep-th",
    reportNumber = "YITP-25-58",
    doi = "10.1103/pg4r-fy8n",
    journal = "Phys. Rev. Lett.",
    volume = "134",
    number = "24",
    pages = "240001",
    year = "2025"
}

@article{Susskind:2014rva,
    author = "Susskind, Leonard",
    title = "{Computational Complexity and Black Hole Horizons}",
    eprint = "1403.5695",
    archivePrefix = "arXiv",
    primaryClass = "hep-th",
    doi = "10.1002/prop.201500092",
    journal = "Fortsch. Phys.",
    volume = "64",
    pages = "24--43",
    year = "2016",
    note = "[Addendum: Fortsch.Phys. 64, 44--48 (2016)]"
}

@article{Stanford:2014jda,
    author = "Stanford, Douglas and Susskind, Leonard",
    title = "{Complexity and Shock Wave Geometries}",
    eprint = "1406.2678",
    archivePrefix = "arXiv",
    primaryClass = "hep-th",
    doi = "10.1103/PhysRevD.90.126007",
    journal = "Phys. Rev. D",
    volume = "90",
    number = "12",
    pages = "126007",
    year = "2014"
}

@article{Das:2025fcd,
    author = "Das, Rathindra Nath and Kundu, Arnab and Martins Costa, Matheus H. and Sarkar, Nemai Chandra",
    title = "{Temporal correlations and chaos from spacetime kernels}",
    eprint = "2512.06078",
    archivePrefix = "arXiv",
    primaryClass = "hep-th",
    doi = "10.1007/JHEP04(2026)141",
    journal = "JHEP",
    volume = "04",
    pages = "141",
    year = "2026"
}

@article{Das:2026ifj,
    author = "Das, Rathindra Nath and Kundu, Arnab and Sarkar, Nemai Chandra",
    title = "{Timelike Entanglement Signatures of Ergodicity and Spectral Chaos}",
    eprint = "2601.19981",
    archivePrefix = "arXiv",
    primaryClass = "hep-th",
    month = "1",
    year = "2026"
}

@article{Jena:2024tly,
    author = "Jena, Siddhi Swarupa and Mahapatra, Subhash",
    title = "{A note on the holographic time-like entanglement entropy in Lifshitz theory}",
    eprint = "2410.00384",
    archivePrefix = "arXiv",
    primaryClass = "hep-th",
    doi = "10.1007/JHEP01(2025)055",
    journal = "JHEP",
    volume = "01",
    pages = "055",
    year = "2025"
}

@article{Giataganas:2025ize,
    author = "Giataganas, Dimitrios",
    title = "{Timelike Entanglement Entropy and Renormalization Group Flow Irreversibility}",
    eprint = "2512.16499",
    archivePrefix = "arXiv",
    primaryClass = "hep-th",
    month = "12",
    year = "2025"
}

@article{Giataganas:2025div,
    author = "Giataganas, Dimitrios",
    title = "{Holographic Timelike c-function}",
    eprint = "2505.20459",
    archivePrefix = "arXiv",
    primaryClass = "hep-th",
    month = "5",
    year = "2025"
}

\end{document}